 \documentclass[twocolumn]{aastex62}
\usepackage{enumitem}
\usepackage{hhline}
\shorttitle{Superluminous Supernova Light Curve Shapes}
\shortauthors{Chatzopoulos \& Tuminello}

\begin{document}

\title{A SYSTEMATIC STUDY OF SUPERLUMINOUS SUPERNOVA LIGHTCURVE MODELS USING CLUSTERING}

\correspondingauthor{Emmanouil Chatzopoulos}
\email{chatzopoulos@phys.lsu.edu}

\author[0000-0002-0786-7307]{E. Chatzopoulos}
\affiliation{Department of Physics \& Astronomy, Louisiana State University, Baton Rouge, LA, 70803, USA}
\affiliation{Hearne Institute of Theoretical Physics, Louisiana State University, Baton Rouge, LA, 70803, USA}

\author{Richard Tuminello}
\affiliation{Department of Physics \& Astronomy, Louisiana State University, Baton Rouge, LA, 70803, USA}



\begin{abstract}

Superluminous supernova (SLSN) lightcurves exhibit a superior diversity compared to their regular luminosity counterparts in terms of
rise and decline timescales, peak luminosities and overall shapes. It remains unclear whether this striking variety arises due to
a dominant power input mechanism involving many underlying parameters, or due to contributions by different progenitor channels. 
In this work, we propose that a systematic quantitative study of SLSN lightcurve timescales and shape properties, such as symmetry
around peak luminosity, can be used to characterize these enthralling stellar explosions. We find that applying clustering analysis
on the properties of model SLSN lightcurves, powered by either a magnetar spin--down or a supernova ejecta--circumstellar interaction 
mechanism, can yield a distinction between the two, especially in terms of lightcurve symmetry. We show that most events in
the observed SLSN sample with well--constrained lightcurves and early detections strongly associate with clusters
dominated by circumstellar interaction models. Magnetar spin--down models also show association at a lower--degree but 
have difficulty in reproducing fast--evolving and fully symmetric lightcurves. We believe this is due to the truncated
nature of the circumstellar interaction shock energy input as compared to decreasing but continuous power input sources like
magnetar spin--down and radioactive $^{56}$Ni decay. 
Our study demonstrates the importance of clustering analysis in characterizing SLSNe based on high--cadence photometric 
observations that will be made available in the near future by surveys like {\it LSST}, {\it ZTF} and {\it Pan--STARRS}.

\end{abstract}

\keywords{(stars:) supernovae: general -- (stars:) circumstellar matter -- stars: magnetars -- methods: data analysis}


\section{Introduction} \label{intro}

Superluminous supernovae (SLSNe; \citealt{2012Sci...337..927G,2018arXiv181201428G,2018SSRv..214...59M})
possess a striking diversity in terms of photometric and spectroscopic properties. SLSNe are often
divided in two classes based on the presence of hydrogen (H) in their spectra: H--poor (SLSN--I)
and H--rich (SLSN--II) events. In terms of photometry, SLSNe are characterized by reaching
very high peak luminosities ($\gtrapprox 10^{44}$~erg~s$^{-1}$) over timescales ranging from a 
few days to several months. The overall evolution and shape of SLSN lightcurves (LCs) can significantly vary
from one event to another. Some SLSN LCs appear to have a symmetric, bell--like shape around peak 
luminosity \citep{2009ApJ...690.1358B,2011Natur.474..487Q} 
while others are highly skewed with a fast rise followed by a slow, long--term decline \citep{2011ApJ...735..106D,2016ApJ...831..144L}.
Most SLSNe appear to be hosted in low--metallicity dwarf galaxies similar to long--duration
Gamma--ray bursts (LGRBs) \citep{2011ApJ...727...15N,2014ApJ...787..138L}.

Several power input mechanisms have been proposed to interpret the extreme peak luminosities 
and diverse observational properties of SLSNe. Most SLSN--II show robust signs of circumstellar
interaction with a hydrogen medium in their spectra indicating that effective conversion of
shock heating to luminosity can reproduce their LCs \citep{2007ApJ...671L..17S,2013ApJ...773...76C}. 
SLSN--I, on the other hand, do not show the usual signatures of circumstellar interaction and are 
often modelled by magneto--rotational energy release due to the spin--down of a newly--born magnetar following a
core--collapse supernova (CCSN) explosion \citep{2010ApJ...717..245K,2010ApJ...719L.204W,2013ApJ...770..128I}.

Nonetheless, the association between power input mechanism and SLSN type is still ambiguous.
The magnetar spin--down model is occasionally invoked as an explanation to SLSN--II that exhibit 
P--Cygni H$\alpha$ line profiles, like SN~2008es, \citep{2010ApJ...717..245K,2018A&A...610L..10D}. 
On the other hand, circumstellar interaction cannot be completely ruled out for SLSN--I events because H 
lines may be hidden due to complicated circumstellar matter geometries \citep{2018ApJ...856...29M,2018MNRAS.475.3152K},
details of non--local thermal equillibrium line transfer physics in non--homologously expanding 
shocked, dense regions yet unexplored by numerical radiation transport models 
\citep{2013ApJ...773...76C,2015MNRAS.449.4304D} or, simply, interaction with a 
H--deficient medium \citep{2012ApJ...760..154C,2016ApJ...828...94C,2016ApJ...829...17S}. A sub--class of SLSNe
are found to transition from SLSN--I at early times to SLSN--II of Type IIn at late times indicating
late--time interaction adding to the complexity of the problem \citep{2017ApJ...848....6Y}.

Breaking the degeneracy between SLSNe powered by magnetar spin--down, circumstellar interaction
and other mechanisms will help address a variety of important questions surrounding massive
stellar evolution and explosive stellar death: the link between LGRBs and SLSNe, the formation
of extremely magnetized stars following CCSN and their effect on the dynamics of the expansion
of the supernova (SN) ejecta, the mass--loss history of massive stars in the days to years prior to their explosion
and how their environments affect the radiative properties of their explosion, to name a few.

The advent of automated, wide--field, high--cadence transient surveys like the 
{\it Panoramic Survey Telescope and Rapid Response System; Pan--STARRS} \citep{2002SPIE.4836..154K}, 
the {\it Zwicky Transient Facility; ZTF} \citep{2019PASP..131a8002B} and, of course, the {\it Large Synoptic Survey
Telescope (LSST)} \citep{2008SerAJ.176....1I} will significantly enhance the SLSN discovery rate and equip
us with more complete photometric coverage that includes detections shortly after the SN explosion 
tightly constraining the LCs of these events. 

This work aims to illustrate how well--sampled LCs can be used to unveil the power input mechanism of
SLSNe. This is done by quantitatively characterizing several key properties of SLSN LCs such as rise and decline
timescales \citep{2015MNRAS.452.3869N} and LC symmetry around peak luminosity. 
Using the power of machine learning and $k$--means clustering analysis we are able to distinguish 
between groups of LC shape parameters corresponding to different power input mechanisms, and 
calculate their association with the properties of observed SLSN LCs. 

Our paper is organized as follows: Section~\ref{obs} presents the observed SLSN LC sample
that we use in this work and introduces the LC shape properties that are utilized in our
analysis. Section~\ref{mod} introduces the SLSN power input models adopted
to obtain large grids of semi--analytic LCs across the associated parameter spaces.
Section~\ref{cluster} introduces the $k$--means clustering analysis method that we employ
to characterize observed and model SLSN LCs and Section~\ref{results} details the results of
this analysis. Finally, Section~\ref{disc} summarizes our discussion.

\section{Observed SLSN Lightcurve Sample}\label{obs}

We use the {\it Open Supernova Catalog} (OSC; \citealt{2017ApJ...835...64G}) to access publicly available
photometric data on a sample of 126 events that are spectroscopically classified as SLSN--I (68\% of the sample) or SLSN--II (32\% of the sample).

For events with available redshift measurements, we compute pseudo--bolometric LCs using the {\it SuperBol}\footnote{https://github.com/mnicholl/superbol} 
code \citep{2018RNAAS...2d.230N}. {\it SuperBol} is a user--friendly {\it Python} software instrument that uses the available observed fluxes in different filters
to fit blackbodies to the Spectral Energy Distribution (SED) of a SN. The resulting pseudo--bolometric SN LCs can also be corrected for time dilation, distance and converted to the
rest frame (K--correction). Using extrapolation techniques, missing near--infrared (NIR) and ultraviolet (UV) flux can also be accounted for. 
Subsequently, all rest--frame LCs are translated in time so that $t =$~0 is coincident with the time corresponding to peak luminosity ($t_{\rm 0} = t_{\rm max}$), 
and scaled by the peak luminosity ($L_{\rm max}$).

For the purposes of our study, we select a sub--sample of SLSNe defined by rest--frame LCs with near--complete temporal photometric coverage,
that we define as including observed data in the range $L_{\rm max}/{\rm e}<L(t)<L_{\rm max}$ (or $1/{\rm e} < L(t) < 1$ in the scaled form). 
Thus, we only focus on SLSN LCs with observed evolution within one ${\rm e}$--folding timescale
from the peak luminosity ensuring that our analysis relies only on real data and not approximate, often model--based,
extrapolations to explosion time (see~\ref{lcshape}). 
In this regard, our sample selection criterion for LC coverage is similar to that used in 
(\citealt{2015MNRAS.452.3869N}; hereafter referred to as N15) 
but our SLSN sample is larger from their ``gold'' sample by 8 events due to our inclusion of SLSN--II events and the availability of 
more SLSN discoveries since their publication.
This process leaves us with a reduced sample of 25 SLSNe with well--covered LCs: 21 SLSN--I and 4 SLSN--II events. 
Table~\ref{T1} presents the details of the SLSN sample used in our analysis including the photometric band 
with the longest (in time) LC coverage that was used in generating their pseudo--bolometric LC.

\subsection{{\it Quantitative properties of SLSNe LC shapes}}\label{lcshape}

In order to quantitatively constrain the shapes of SLSN LCs, we define the following scaled luminosity
thresholds:
\begin{itemize}
\item{Primary luminosity threshold: $L_{\rm 1} =1.0/{\rm e}$ or 36.79\% of the peak luminosity.}
\item{Secondary luminosity threshold: $L_{\rm 2} = 1.0/(0.5{\rm e})$ or 73.58\% of the peak luminosity.}
\item{Tertiary luminosity threshold: $L_{\rm 3} = 1.0/(0.4{\rm e})$ or 91.97\% of the peak luminosity.}
\end{itemize}
At each luminosity threshold we can compute a ``rise--time'' to peak luminosity and a ``decline--time''
from peak. As such, we accordingly define the primary, secondary and tertiary rise ($tr_{\rm 1}$, $tr_{\rm 2}$, $tr_{\rm 3}$) 
and decline ($td_{\rm 1}$, $td_{\rm 2}$, $td_{\rm 3}$) timescales. It is evident that $t[d,r]_{\rm 3} <  t[d,r]_{\rm 2} < t[d,r]_{\rm 1}$
and that all of the SLSNe in our selected LC sample have observations that include these timescales.
We note that our choice for the primary luminosity threshold and corresponding rise and decline timescales
is the same as the one used in N15 to study how closely these timescales
correlate with different power input models.

Next, for the sake of quantifying how symmetric a LC is around peak luminosity, we define three corresponding
``LC symmetry'' parameters: $s_{\rm 1,2,3} = tr_{\rm 1,2,3}/td_{\rm 1,2,3}$. The closer these parameters
are to unity, the more symmetric the LC is at the corresponding luminosity threshold. Obviously, to
consider a LC as ``fully symmetric'' all of the three LC symmetry parameters need to be close to unity. For
the purposes of this study we define a symmetric LC one that satisfies the criterion $0.9 < s_{\rm 1,2,3} < 1.1$.
For the remainder of this paper we refer to the nine ($tr_{\rm 1,2,3}$, $td_{\rm 1,2,3}$, $s_{\rm 1,2,3}$) LC parameters
as ``LC shape parameters''.

\begin{figure}
\begin{center}
\includegraphics[angle=0,width=9cm,trim=0.5in 0.25in 0.5in 0.15in,clip]{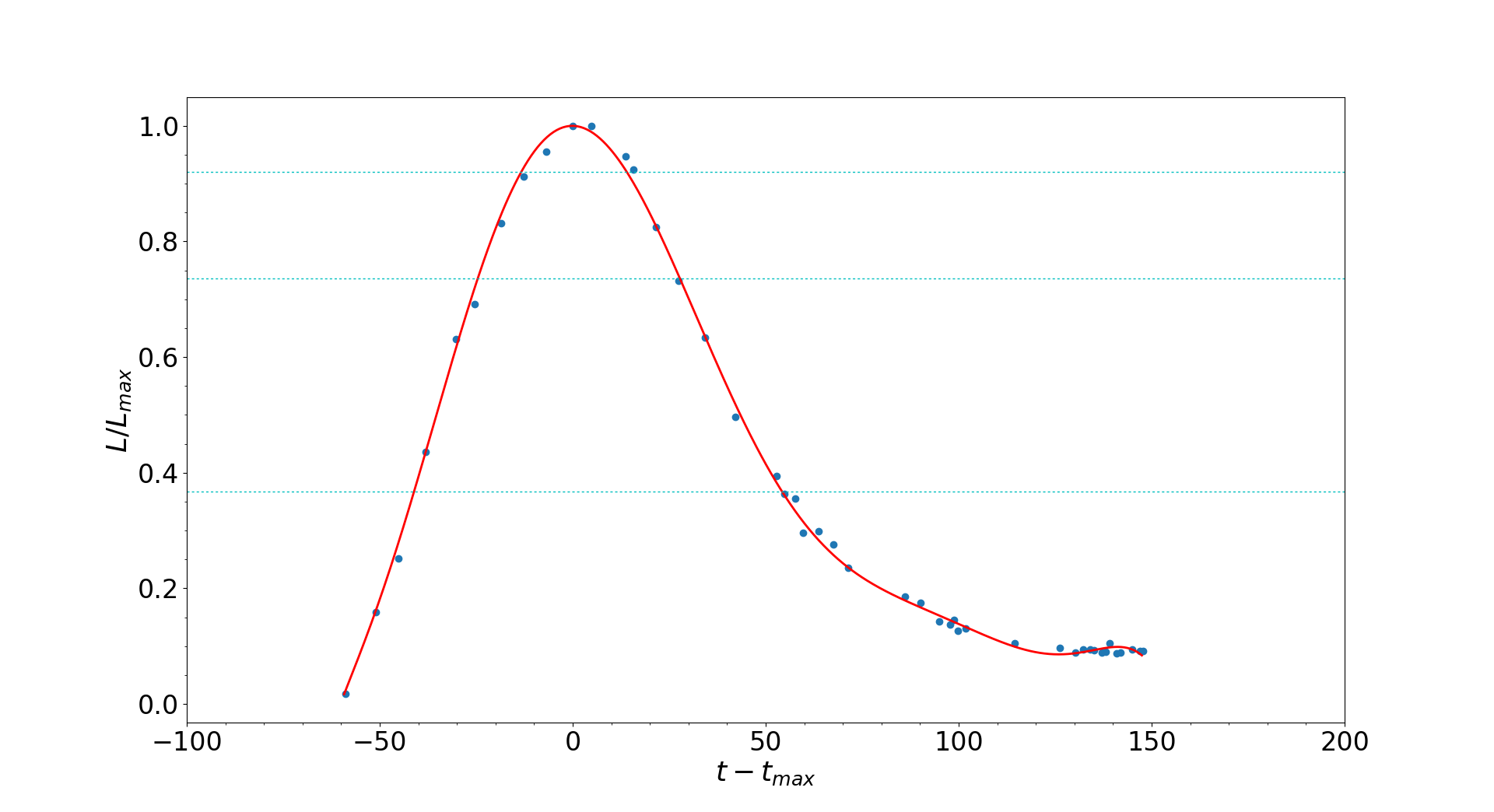}
\caption{Polynomial fit (red curve) to the observed scaled pseudo--bolometric LC of Type II SN2006gy (blue circles). An 8th--degree polynomial is used for the fit.
The primary, secondary and tertiary luminosity thresholds are shown as horizontal light blue lines.}
\label{Fig:06gypoly}
\end{center}
\end{figure}

\setcounter{table}{0}
\begin{deluxetable*}{lccccccccccccc}
\tabletypesize{\footnotesize}
\tablecaption{The SLSN LC sample used for this work.}
\tablehead{
\colhead {SLSN} &
\colhead{Reference} &
\colhead {$z$} &
\colhead{Filters$^{\dagger}$} &
\colhead {$tr_{\rm 1}$} &
\colhead {$td_{\rm 1}$} &
\colhead {$s_{\rm 1}$} &
\colhead {$tr_{\rm 2}$} &
\colhead {$td_{\rm 2}$} &
\colhead {$s_{\rm 2}$} &
\colhead {$tr_{\rm 3}$} &
\colhead {$td_{\rm 3}$} &
\colhead {$s_{\rm 3}$} &
\\}
\startdata
& & &  {\it SLSN--I} & & & & & & & & & \\
\hline
\hline
PTF09cnd & \citet{2011Natur.474..487Q} & 0.258 & UBgRi & 29.5 & 56.3 & 0.52 & 18.9 & 26.9 & 0.7 & 10.6 & 12.9 & 0.82 \\ 
SN2011kg & \citet{2013ApJ...770..128I} & 0.192 & UBgrizJ & 20.5 & 30.0 & 0.68 & 12.5 & 15.9 & 0.79 & 6.9 & 7.9 & 0.88 \\ 
SN2010md & \citet{2013ApJ...770..128I} & 0.098 & UBgriz & 30.4 & 31.9 & 0.95 & 16.1 & 16.6 & 0.97 & 8.4 & 8.4 & 1.0 \\ 
SN2213--1745 & \citet{2012Natur.491..228C} & 2.046 & g$^{\prime}$r$^{\prime}$i$^{\prime}$ & 10.4 & 25.5 & 0.41 & 6.7 & 8.6 & 0.78 & 3.7 & 4.3 & 0.87 \\
PTF09atu & \citet{2011Natur.474..487Q} & 0.501 & gRi & 48.8 & 50.9 & 0.96 & 29.9 & 30.2 & 0.99 & 16.4 & 16.0 & 1.02 \\
iPTF13ajg & \citet{2014ApJ...797...24V} & 0.740 & uBgR$_{\rm s}$iz & 21.9 & 28.8 & 0.76 & 14.3 & 16.4 & 0.87 & 8.0 & 8.6 & 0.93 \\ 
PS1--10pm & \citet{2015MNRAS.448.1206M} & 1.206 & griz & 27.9 & 25.4 & 1.1 & 14.9 & 15.0 & 0.99 & 7.9 & 7.9 & 1.0 \\
PS1--14bj & \citet{2016ApJ...831..144L} & 0.522 & grizJ & 81.6 & 138.2 & 0.59 & 49.2 & 64.9 & 0.76 & 27.2 & 32.4 & 0.84 \\ 
SN2013dg & \citet{2014MNRAS.444.2096N} & 0.265 & griz & 15.6 & 29.7 & 0.52 & 10.4 & 14.0 & 0.74 & 5.9 & 6.8 & 0.87 \\
iPTF13ehe & \citet{2015ApJ...814..108Y,2017ApJ...848....6Y} & 0.343 & gri & 53.4 & 62.1 & 0.86 & 32.2 & 35.4 & 0.91 & 18.1 & 18.1 & 1.0 \\
LSQ14mo & \citet{2015ApJ...815L..10L} & 0.253 & Ugri & 16.2 & 25.3 & 0.64 & 10.9 & 14.0 & 0.78 & 6.2 & 7.1 & 0.87 \\ 
PS1--10bzj & \citet{2013ApJ...771...97L} & 0.650 & griz & 14.6 & 22.5 & 0.65 & 10.3 & 13.8 & 0.75 & 6.1 & 7.2 & 0.84 \\ 
DES14X3taz & \citet{2016ApJ...818L...8S} & 0.608 & griz & 31.9 & 41.8 & 0.76 & 19.9 & 23.0 & 0.87 & 11.0 & 11.7 & 0.94 \\ 
LSQ14bdq & \citet{2015ApJ...807L..18N} & 0.345 & griz & 54.6 & 90.2 & 0.61 & 37.1 & 48.8 & 0.76 & 21.7 & 24.4 & 0.89 \\
SNLS 07D2bv & \citet{2013ApJ...779...98H} & 1.500 & griz & 18.9 & 17.7 & 1.07 & 12.5 & 12.8 & 0.98 & 7.1 & 7.0 & 1.01 \\
SNLS 06D4eu & \citet{2013ApJ...779...98H} & 1.588 & griz & 15.0 & 17.6 & 0.85 & 9.4 & 10.6 & 0.89 & 5.3 & 5.7 & 0.92 \\ 
PTF12dam & \citet{2018ApJ...860..100D} & 0.107 & UBgVrizJHK & 46.2 & 75.0 & 0.62 & 28.8 & 37.5 & 0.77 & 16.6 & 18.3 & 0.91 \\
SN2011ke & \citet{2018ApJ...860..100D} & 0.143 & UBgVriz & 22.1 & 26.6 & 0.83 & 12.3 & 13.8 & 0.97 & 6.8 & 7.0 & 0.97 \\
PTF12gty & \citet{2018ApJ...860..100D} & 0.177 & gri & 46.4 & 65.9 & 0.70 & 24.9 & 27.0 & 0.92 & 14.0 & 15.2 & 0.92 \\
PS1--11ap & \citet{2018ApJ...852...81L} & 0.524 & grizy & 26.7 & 52.5 & 0.51 & 18.5 & 26.3 & 0.71 & 11.0 & 12.9 & 0.85 \\
SCP 06F6 & \citet{2009ApJ...690.1358B} & 1.189 & iz & 31.8 & 32.7 & 0.97 & 19.5 & 19.5 & 1.0 & 10.6 & 10.4 & 1.02 \\ 
\hline
& & &  {\it SLSN--II} & & & & & & & & & \\
\hline
SN2006gy & \citet{2007ApJ...666.1116S} & 0.019 & BVR & 41.0 & 54.3 & 0.76 & 24.4 & 27.8 & 0.88 & 13.3 & 14.1 & 0.94 \\ 
CSS121015:004244+132827 & \citet{2014MNRAS.441..289B} & 0.287 & UBVRGI & 20.3 & 30.9 & 0.66 & 12.5 & 15.2 & 0.82 & 7.0 & 7.6 & 0.92 \\ 
SN2016jhn & \citet{2018arXiv180108240M} & 1.965 & GI2zY & 12.4 & 27.0 & 0.46 & 10.3 & 20.7 & 0.5 & 6.3 & 10.6 & 0.6 \\ 
SDSSII SN2538 &\citet{2018PASP..130f4002S} & 0.530 & u$^{\prime}$g$^{\prime}$r$^{\prime}$i$^{\prime}$z$^{\prime}$ & 31.6 & 37.8 & 0.84 & 19.0 & 19.2 & 0.99 & 10.0 & 10.0 & 1.0 \\
\enddata 
\tablecomments{The SLSN LC data were collected from the {\it Open Supernova Catalog} \citep{2017ApJ...835...64G} database. All timescales
are in units of days.
$^{\dagger}$This column lists the filters used to compile the pseudo--bolometric LC of each SLSN in our sample.
\label{T1}}
\end{deluxetable*}

\setcounter{table}{1}
\begin{deluxetable*}{lccccc|cccccccc}
\tabletypesize{\footnotesize}
\tablecaption{Main statistical properties of the observed SLSN LC sample used in this work.}
\tablehead{
\colhead {Parameter} &
\colhead{$\mu$} &
\colhead{$M$} &
\colhead {$\sigma$} &
\colhead{max} &
\colhead {min} &
\colhead{$\mu$} &
\colhead{$M$} &
\colhead {$\sigma$} &
\colhead{max} &
\colhead {min} &
\\}
\startdata
&&& SLSN--I & & & & & SLSN--II & & \\
\hline
\hline
$tr_{\rm 1}$ & 31.6 & 27.9 & 17.3 & 81.6 & 10.4 & 26.3 & 25.9 & 10.9 & 41.0 & 12.4 \\
$td_{\rm 1}$ & 45.1 & 31.9 & 28.5 & 138.2 & 17.6 & 37.5 & 34.3 & 10.4 & 54.3 & 27.0 \\
$s_{\rm 1}$ & 0.74 & 0.70 & 0.19 & 1.10 & 0.41 & 0.68 & 0.71 & 0.14 & 0.84 & 0.46 \\
$tr_{\rm 2}$ & 19.5 & 16.1 & 10.5 & 49.2 & 6.7 & 16.6 & 15.8 & 5.6 & 24.4 & 10.3 \\
$td_{\rm 2}$& 23.4 & 16.6 & 13.6 & 64.9 & 8.6  & 20.7 & 19.9 & 4.5 & 27.8 & 15.3 \\
$s_{\rm 2}$ & 0.85 & 0.87 & 0.10 &  1.00 & 0.70 & 0.80 & 0.85 & 0.18 & 0.99 & 0.50 \\
$tr_{\rm 3}$ & 10.9 & 8.4 & 5.9 & 27.2 & 3.7 & 9.1 & 8.5 & 2.8 & 13.3 & 6.2 \\
$td_{\rm 3}$ & 11.9 & 8.6 & 6.7 & 32.4 & 4.3 & 10.6 & 10.3 & 2.3 & 14.1 & 7.6 \\
$s_{\rm 3}$ & 0.93 & 0.92 & 0.06 & 1.02 & 0.82 & 0.86 & 0.93 & 0.16 & 1.00 & 0.60 \\
\hline
\enddata 
\tablecomments{The parameters $\mu$, $M$, $\sigma$, max, and min correspond to the values of the mean, median, standard deviation,
maximum and minimum of the sample accordingly. All timescales are in units of days.
\label{T2}}
\end{deluxetable*}

We have developed a {\it Python} script that fits a high--degree polynomial to the scaled observed LCs of
the SLSN in our sample. This provides with interpolation between missing photometric datapoints and an accurate
measurement of the LC shape parameters discussed above. An example of such fit is shown in Figure~\ref{Fig:06gypoly}
for SN2006; unarguably one of the most well--observed SLSN--II of Type IIn \citep{2007ApJ...666.1116S}. 
In this figure, the light blue horizontal lines
show the three luminosity thresholds that were introduced earlier. Based on these thresholds, we find
$tr_{\rm 1} =$~41.0~days and $td_{\rm 1} =$~54.3~days for this SN, implying primary symmetry, $s_{\rm 1} =$~0.76.
The rest of the LC shape parameters for SN2006gy are given in Table~\ref{T1}. 
Table~\ref{T2} lists the main LC shape statistical properties of the observed SLSN--I and SLSN--II in our sample. The SLSN--II 
sample only includes 4 events therefore preventing us from performing an accurate statistical comparison against
the SLSN--I sample to look for potential systematic differences in the two distributions. 

Our sample overlaps with that presented in Table 3 of N15 for 11 SLSNe: SN2011ke, SN2013dg, LSQ14mo, LSQ13bdq, PTF12dam, 
CSS121015:004244+132827, PS1--11ap, SCP 06F6, PTF09cnd, PS1--10bj and iPTF13ajg. 
This is due to the fact that for the purposes of our study we decided to include only
events with real detections shortly after the explosion and a good coverage of the LC in order to tightly constrain their LC shape parameters.
N15, on the other hand, opted to use polynomial extrapolation to earlier times for some of the SLSNe in
their sample in order to obtain estimates for $tr_{\rm 1}$ and $td_{\rm 1}$. For objects where this extrapolation is done only by a few
days this may not be a bad approximation, however the LCs for cases like SN2007bi \citep{2009Natur.462..624G}, SN2005ap 
\citep{2007ApJ...668L..99Q}, and PS1-10ky \citep{2011ApJ...743..114C}, $tr_{\rm 1}$ is poorly constrained using this method.

For the 11 events that are common between our sample and that of N15, we calculate the mean
value of $tr_{\rm 1}$ to be 27.2~days versus 25.7~days in their case, and the mean value of $td_{\rm1}$ to be 42.8~days compared
to 51.6~days in their case. While our results are consistent in terms of $tr_{\rm 1}$, the discrepancy observed in $td_{\rm 1}$ could
be due to a variety of reasons including different combinations of filters used to calculate the rest--frame pseudo--bolometric LC
of each event. In our work, we have used all available filters with more than 2 data points for each event to construct LCs using
{\it SuperBol} as described earlier. We caution that more accurate consideration for near--IR and IR fluxes may lead to flattening of the
true bolometric LC at late times and therefore longer primary decline timescales.

We note that comparing the mean $tr_{\rm 1}$ and $td_{\rm 1}$ values of our entire sample 
($tr_{\rm 1} =$~30.8~days, $td_{\rm 1} =$~43.9~days from Table~\ref{T2}) 
against those of the full SLSN sample of N15 (their Table 3; $tr_{\rm 1} =$~22.9~days, $td_{\rm 1} =$~46.4~days) the agreement
is somewhat better, within uncertainties. We also derive a linear fit for the observed $tr_{\rm 1}$ and $td_{\rm 1}$ values of the form:
\begin{equation}
td_{\rm 1} =  \gamma_{\rm 0} + \gamma_{\rm 1} \times tr_{\rm 1}\label{Eq1},
\end{equation}
where $\gamma_{\rm 0} =$~-1.962 and $\gamma_{\rm 1} =$~1.489 (see also~\ref{Fig:tr1td1fit}). In contrast, N15
derive a steeper correlation for their ``gold'' SLSN sample with $\gamma_{\rm 0,N15} =$~-0.10 and $\gamma_{\rm 1,N15} =$~1.96.

An investigation of Table~\ref{T1} reveals yet another interesting property of our observed SLSN sample: five SLSN--I events
SN2010md, PTF09atu, PS1--10pm, SNLS 07D2bv and SCP 06F6) or, equivalently, 23.81\% of the entire SLSN--I sample have
fully symmetric LC around peak luminosity, following the criterion we established earlier for full LC symmetry ($0.9 < s_{\rm 1,2,3} < 1.1$).
This can be said for more certainty for SN2010md and PTF09atu (with redshifts 0.098 and 0.5 accordingly) as compared
to the other three events with large redshifts ($>$~1), because in this case the observed band correspond to near--UV fluxes
in the rest--frame. Bias toward UV fluxes may correspond to faster post--maximum decline rate and thus steeper, more symmetric
LCs. Neverthelss, we have attempted to account for this effect by making use of approximate extrapolations to the IR flux by using the
techniques available in {\it SuperBol}.

The upper left panel of Figure~\ref{Fig:symmLCs} shows 2 examples of SLSNe with ``fully--symmetric'' LCs. 
Given that symmetric LCs are present in about a quarter of our SLSN--I sample, a considerable fraction of LC models corresponding to the 
proposed power input mechanisms must be able to reproduce this observation. {\it This raises the question of whether
LC symmetry is a property shared amongst all the proposed power input mechanisms for different combinations of model parameters
or is uniquely tied to one power input mechanism. In the latter case, we can use photometry alone to characterize
the nature of SLSNe}.

Lastly, another LC shape property that will be interesting to constrain with future, high--cadence photometric follow--up of SLSNe would
be the convexity (second derivative) of the bolometric LC during the rise to peak luminosity \citep{2017ApJ...851L..14W}. Given the
low temporal resolution of the observed LC in our sample, we opt to not provide estimates of the percentages of concave--up and concave--down
LCs, yet we briefly discuss the predictions for these parameters coming from semi--analytical models in the following section.

\section{SLSN Power Input Models}\label{mod}

A number of models have been proposed to explain both the unprecedented peak luminosities
but, more importantly, the striking diversity in the observed properties of SLSNe, both photometrically (LC timescales and shapes) 
and spectroscopically (SLSN--I versus SLSN--II class events). The three most commonly cited SLSN power input mechanisms are
the radioactive decay of several masses of $^{56}$Ni produced in a full--fledged Pair--Instability Supernova explosion
(PISN; \citealt{2009Natur.462..624G,2012ApJ...748...42C,2015ApJ...799...18C}), the magneto--rotational energy release from
the spin--down of a newly born magnetar following a core--collapse SN (CCSNe) \citep{2010ApJ...717..245K,2010ApJ...719L.204W}
and the interaction between SN ejecta and massive, dense circumstellar shells ejected by the progenitor star prior to the explosion
\citep{2007ApJ...671L..17S,2008ApJ...686..467S,2016ApJ...828...94C,2017ApJ...851L..14W}. 

We have decided to leave the PISN model outside of our analysis because of several reasons that make it unsuitable for
contemporary SLSNe. First, given that the known hosts of SLSNe have metallicities $Z >$~0.1 \citep{2013ApJ...771...97L,2014ApJ...787..138L},
very massive stars formed in these environments are likely to suffer strong radiatively--driven mass--loss preventing them
from forming the massive carbon--oxygen cores 
($\gtrapprox$~40--60~$M_{\odot}$, depending on Zero Age Main Sequence rotation rate \citealt{2012ApJ...748...42C})
required to encounter pair--instability \citep{2007A&A...475L..19L}. 
Second, the majority of PISN models do not yield superluminous LCs. Yet even many of the PISN superluminous LCs require
total SN ejecta masses that are comparable to -- or smaller in some cases -- to the predicted
$^{56}$Ni mass needed to explain the high peak luminosity \citep{2013ApJ...773...76C}. 
Finally, while radiation transport models of PISNe can reproduce superluminous LCs and provide good fits to the LCs of some 
SLSNe \citep{2009Natur.462..624G,2017ApJ...846..100G},
the model spectra are too red compared to the observed SLSN spectra at contemporaneous epochs \citep{2013MNRAS.428.3227D,2015ApJ...799...18C}.
Full--fledged PISN may however still be at play in lower metallicity environments and massive, Population III primordial stars. For an alternative
perspective on the viability of low--redshfit full--fledged PISNe we refer to \citet{2014A&A...565A..70K}.

We add that a model that is recently gaining popularity is energy input by fallback accretion into a newly--formed black hole
following core collapse \citep{2013ApJ...772...30D}. One caviat of this model is that unrealistically large accretion masses are needed
in order to fit the observed LCs of SLSNe given a fiducial choice for the energy conversion efficiency for the most cases \citep{2018ApJ...867..113M}. 
While the fallback accretion model is a very interesting suggestion that may be relevant to a small fraction of SLSNe, we opt to exclude it from our
model LC shape analysis at least until it is further investigated in the literature. This leaves us with two main channels to power SLSNe often
discussed today, the magnetar spin--down and the cirumstellar interaction model. From hereafter, we refer to the magnetar spin--down model as 
``MAG'' and to the SN ejecta--circumstellar interaction model as ``CSM''.

For both the MAG and the CSM model, we adopt the semi--analytic formalism presented in \citep{2012ApJ...746..121C,2013ApJ...773...76C}
(hereafter C12, C13) and based on the seminal works of 
\citep{1980ApJ...237..541A,1982ApJ...253..785A} on modeling the LCs of Type Ia and Type II SNe.
While these models invoke many simplifying assumptions (centrally concentrated input source -- in terms of energy density, homologous
expansion of the SN ejecta and constant, Thompson scattering opacity for the SN ejecta to name a few), they remain a powerful tool
to study the LC shapes of SNe assuming different power inputs because of their ability to provide reasonable estimates of the 
associated physical parameters when fit to observed data. In addition, these semi--analytic models are numerically inexpensive to compute,
allowing us to compute large grids of LC models throughout the associated, multi--dimensional parameter space. As such they remain
a popular SN LC modeling tool with a few codes that have been made publicly available to compute them
such as {\it TigerFit} \citep{2017ApJ...851L..14W} and {\it MOSFiT} \citep{2018ApJS..236....6G}. 
We caution, however, that comparisons against rigorous, numerical radiation transport models
have shown that semi--analytic SLSN LC models have their limitations, especially in regimes where the SN expansion is not homologous
(for example due to circumstellar interaction) and due to the assumption of constant opacity in the SN ejecta and constant diffusion timescale 
\citep{2013MNRAS.428.1020M,2018arXiv181206522K}. For this reason, we include some analysis of the LC shape properties of 
numerically--computed SLSN LCs that are available in the literature for both the MAG and the CSM model.

\subsection{{\it The SN--ejecta circumstellar interaction model (CSM)}}\label{csi}

Massive stars can suffer significant mass--loss episodes, especially during the late stages of their evolution, due to a variety
of mechanisms: super--Eddington strong winds during a Luminous Blue Variable (LBV) stage similar to the $\eta$--Carina 
\citep{2007ApJ...671L..17S,2011MNRAS.415..773S,2018arXiv180910187J,2018MNRAS.480.1466S},
gravity--wave driven mass--loss excited during vigorous shell Si and O shell burning 
\citep{2012MNRAS.423L..92Q,2014ApJ...780...96S,2017MNRAS.470.1642F},
binary interactions \citep{1994ApJ...429..300W} or a softer version of PISN that does not lead to complete disruption of
the progenitor star (Pulsational Pair--Instability or PPISN; \citealt{2007Natur.450..390W,2012ApJ...760..154C,2017ApJ...836..244W}).
PPISNe originate from less massive progenitors than full--fledged PISNe and can thus occur in the nearby Universe 
offering a channel to produce a sequence of SLSN--like transients originating from the same progenitor as successively ejected 
shells can collide with each other before the final CCSN takes place \citep{2016ApJ...828...94C,2017ApJ...836..244W,2018NatAs.tmp..125L}.

As a result, both observational evidence and theoretical modeling suggest that the environments around massive stars
can be very complicated with diverse geometries (circumstellar (CS) spherical or bipolar shells, disks or clumps) and, in some cases,
very dense and at the right distance from the progenitor star that a violent interaction will be imminent following the SN explosion.
This SN ejecta--circumstellar matter interaction (CSI) leads to the formation of forward and reverse shocks and the efficient conversion of kinetic
energy into luminosity \citep{1994ApJ...420..268C,2041-8205-729-1-L6} 
that can produce superluminous transients with immense diversity in their LC shapes and maybe even
spectra \citep{2012ApJ...747..118M,2013MNRAS.430.1402M,2016MNRAS.tmp..117D,2018MNRAS.475.3152K}.

C12 combined the self--simular CSI solutions presented by \citet{1994ApJ...420..268C} with the \citet{1980ApJ...237..541A,1982ApJ...253..785A}
LC modeling formalism to compute approximate, semi--analytical CSM models that were then successfully fit to the LCs of several SLSN--I and SLSN--II
events in C13. Given a SN explosion energy ($E_{\rm SN}$), SN ejecta mass ($M_{\rm ej}$), the index of the outer (power--law) density profile
of the SN ejecta ($n$, related to the progenitor radius), the distance of the CS shell ($R_{\rm CS}$), the mass of the CS shell $M_{\rm CS}$, the
(power--law) density profile of the CS shell ($s$) and the progenitor star mass--loss rate ($\dot{M}$) a model, semi--analytic 
CSM LC can be computed. 
The energy input originates from the efficient conversion of the kinetic energy of both the forward and the reverse shock to luminosity.
As such, forward shock energy input is terminated when it breaks out to the optically--thin CS while reverse shock input is terminated once it sweeps--up
the bulk of the SN ejecta. 
{\it This is a property unique to the CSM model and not present in other, continuous heating sources such as radioactive decay of $^{56}$Ni
and magnetar spin--down input: during CSI energy input terminates abruptly, thus affecting the shape of the LC in a way that can yield a
faster decline in luminosity at late times.}

While the CSM model can naturally explain the observed diversity of SLSN LCs and is consistent with observation of narrow emission lines
in the spectra of SLSN--II events of IIn class, it has been challenged as a viable explanation for SLSN--I due to the lack of spectroscopic
signatures associated with interaction (\citealt{2013ApJ...770..128I}, N15). There is, however, a ``hybrid'' class of SLSNe that transition from
SLSN--I to SLSN--II at late times indicating possible interaction with H--poor material early on before the SN ejecta reach 
the ejected H envelope and interact with it producing Balmer emission lines \citep{2015ApJ...814..108Y}. 
Another concern for the CSM model is the necessity to include many parameters in the model that can lead to overfitting observed data and to 
parameter degeneracy issues \citep{2013MNRAS.428.1020M}. Detailed radiation hydrodynamics and radiation transport modeling of 
the CSI process across the relevant parameter space, including in cases of H--poor CSI, is still needed in order to resolve whether SLSN--I 
can be powered by this mechanism.

\subsection{{\it The magnetar spin--down model (MAG)}}\label{mag}

The spin--down of a newly born magnetar following CCSN can release magneto--rotational energy that,
if efficiently thermalized in the expanding SN ejecta, can produce a superluminous display \citep{2010ApJ...717..245K,2010ApJ...719L.204W}.
Given a dipole magnetic field for the magnetar, an initial rotation period of $P_{\rm mag}$ in units of 1~ms and an initial magnetar
magnetic field $B_{\rm 14, mag}$ in units of $10^{14}$~G, the associated SN LC can be computed by making use of Equation 13
of C12. This model LC can also provide estimates for the SN ejecta mass, $M_{\rm ej}$, that is controlled by
the diffusion timescale (Equaton 10 of C12).

Numerical radiation transport simulations of SNe powered by magnetars have yielded additional insights on the efficiency
of this model in powering SLSNe, primarily of the hydrogen--poor (SLSN--I) type
\citep{2012MNRAS.426L..76D,2015MNRAS.454.3311M,dessartaudit,2018A&A...610L..10D}. Some
observational evidence linking the host properties of SLSN--I to those of long--duration Gamma--ray bursts \citep{2014ApJ...787..138L}
and the discovery of double--peaked SLSN LCs, a feature that can be produced by magnetar--driven shock breakout 
\citep{2015ApJ...807L..18N,2016ApJ...821...36K}
seem to strongly suggest that most, if not all, SLSN--I are powered by this mechanism. This is strengthened by the suggestion that
a lot of SLSN LCs can be successfully fit but a semi--analytical MAG LC model \citep{2017ApJ...850...55N,2018ApJ...860..100D}.
There is, however, on--going discussion on whether the MAG model is always efficient in thermalizing the magnetar luminosity
in the SN ejecta or even allowing for the efficient conversion of the magnetar energy to radiated luminosity \citep{2006MNRAS.368.1717B}, 
instead of kinetic energy for the inner ejecta \citep{2016ApJ...821...22W}. Recent, 2D simulations of magnetar--powered SNe appear
to enhance these concerns \citep{2016ApJ...832...73C,2017ApJ...839...85C}.

\subsection{{\it Grids of Models with the TigerFit code}}\label{tigerfit}

We have adapted the {\it TigerFit} code \citep{2016ApJ...828...94C,2017ApJ...851L..14W} to run grids of CSM
and MAG models throughout a large parameter space in order to systematically study the statistical LC shape
properties and determine their association with the observed SLSN sample presented in Section~\ref{obs}.

\begin{figure*}
\gridline{\fig{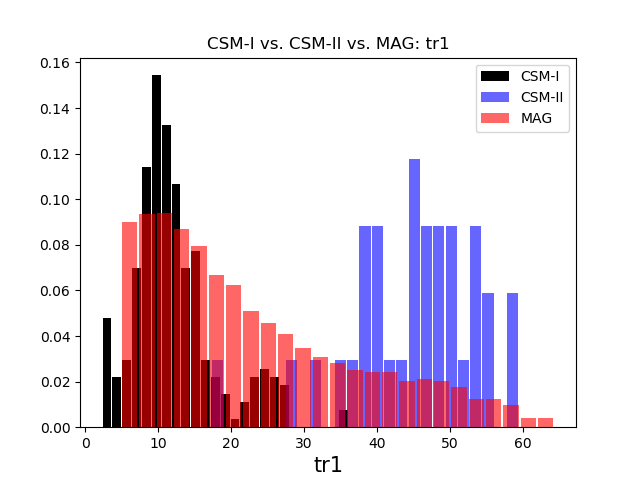}{0.5\textwidth}{}
          \fig{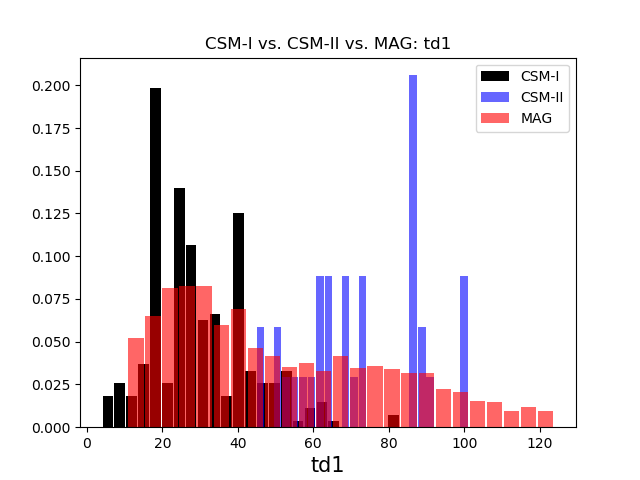}{0.5\textwidth}{}
          }
\caption{Distribution of primary rise ($tr_{\rm 1}$; {\it left panel}) and decline ($td_{\rm 1}$; {\it right panel}) timescales for the CSM--I (black bars),
CSM--II (blue bars) and MAG (red bars) model samples.
\label{Fig:tr1td1}}
\end{figure*}

\begin{figure*}
\gridline{\fig{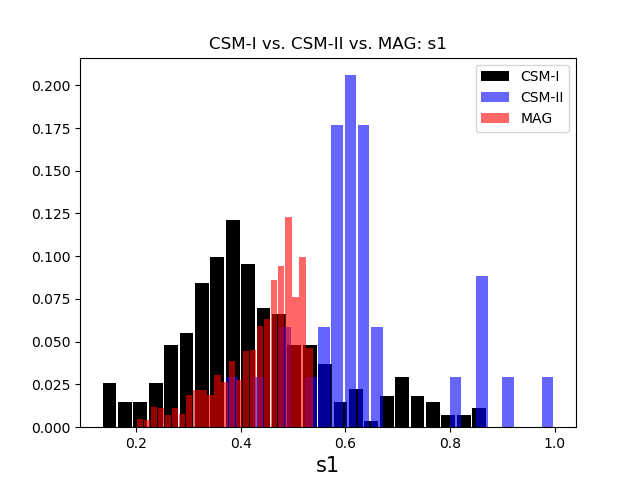}{0.5\textwidth}{}
          \fig{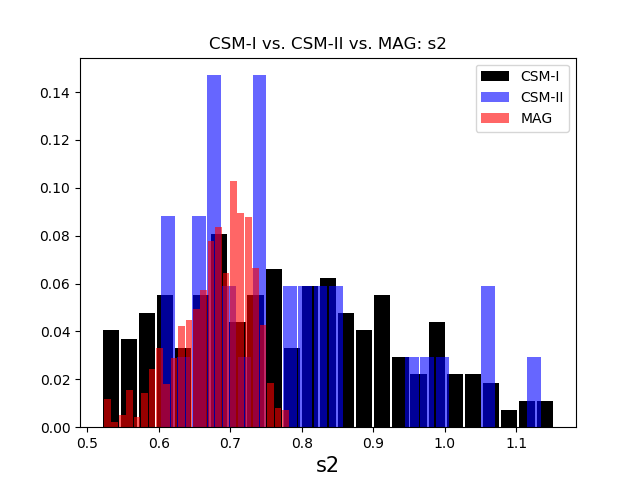}{0.5\textwidth}{}
          }
\gridline{\fig{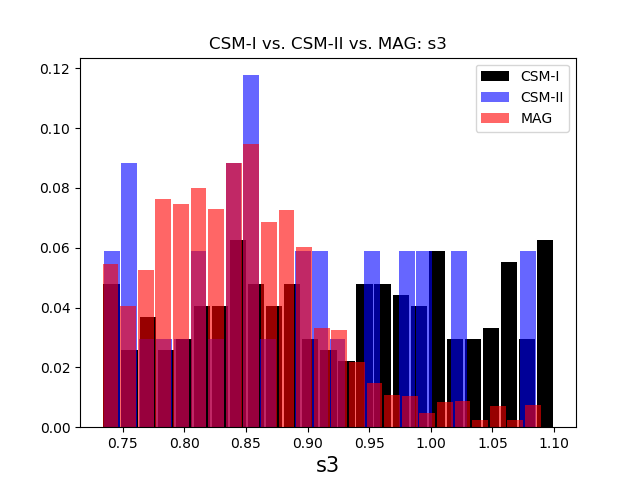}{0.5\textwidth}{}
          }
\caption{Same as Figure~\ref{Fig:tr1td1} but for primary ($s_{\rm 1}$; {\it upper left panel}), secondary ($s_{\rm 2}$; {\it upper right panel}) and 
tertiary ($s_{\rm 3}$; {\it bottom panel}) LC symmetry.
\label{Fig:s1s2s3}}
\end{figure*}

\begin{figure*}
\includegraphics[angle=0,width=18cm,trim=0.5in 0.25in 0.5in 0.5in,clip]{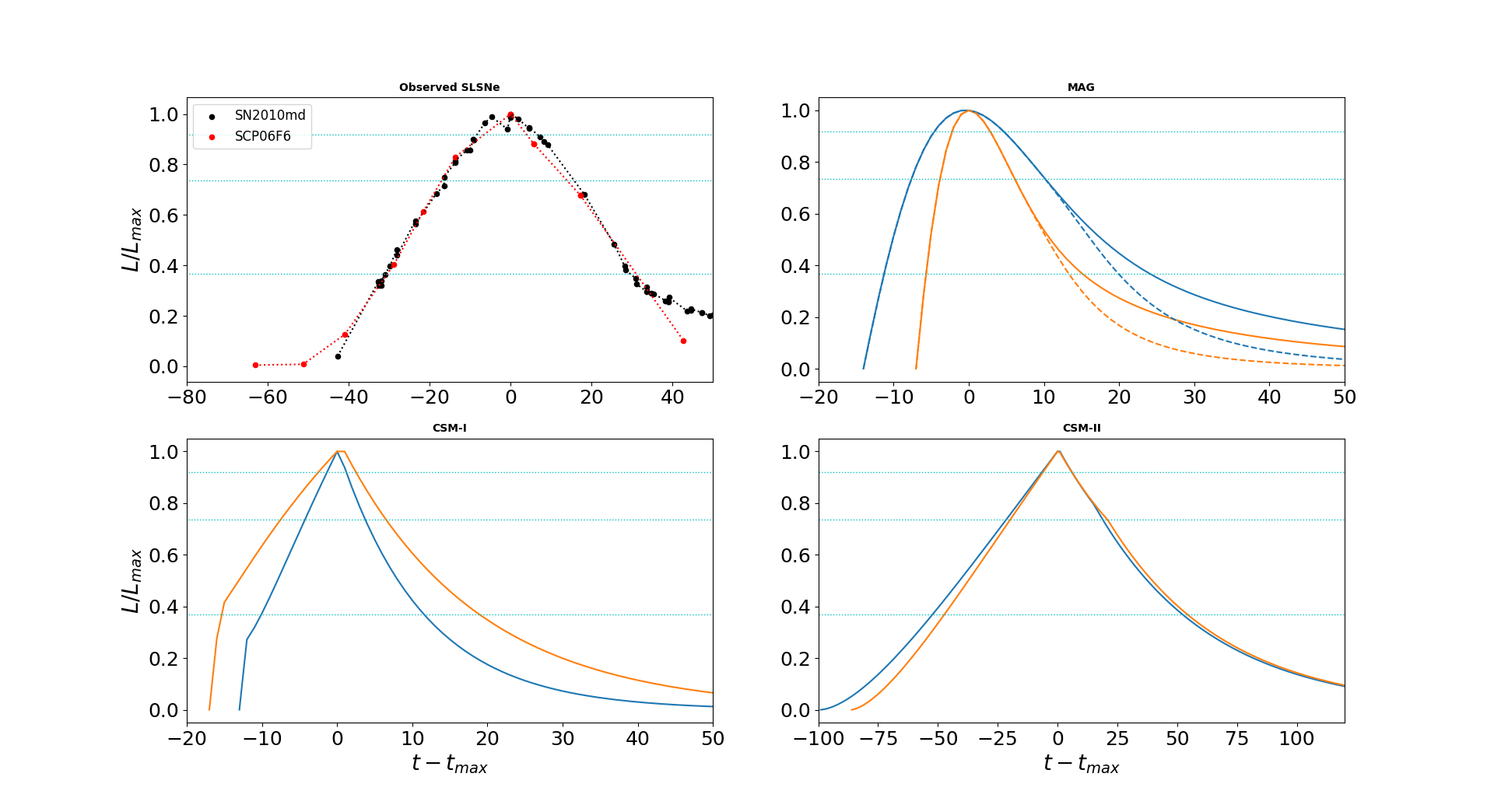}
\caption{The most symmetric LCs of the observed SLSN sample ({\it upper left panel}), the MAG model sample ({\it upper right panel})
the CSM--I model sample ({\it lower left panel}) and the CSM--I model sample ({\it lower right panel}). 
The light blue dashed lines indicate the primary, secondary and tertiary luminosity thresholds used to determine symmetry
around peak luminosity (see~\ref{lcshape}).
It is can be seen that even the most symmetric MAG model LCs are still quite asymmetric at the primary luminosity threshold.
This holds even under the assumption of strong $\gamma$--ray leakage (marked by dashed curves for each MAG model).}
\label{Fig:symmLCs}
\end{figure*}

\begin{figure}
\begin{center}
\includegraphics[angle=0,width=9cm,trim=0.5in 0.25in 0.5in 0.15in,clip]{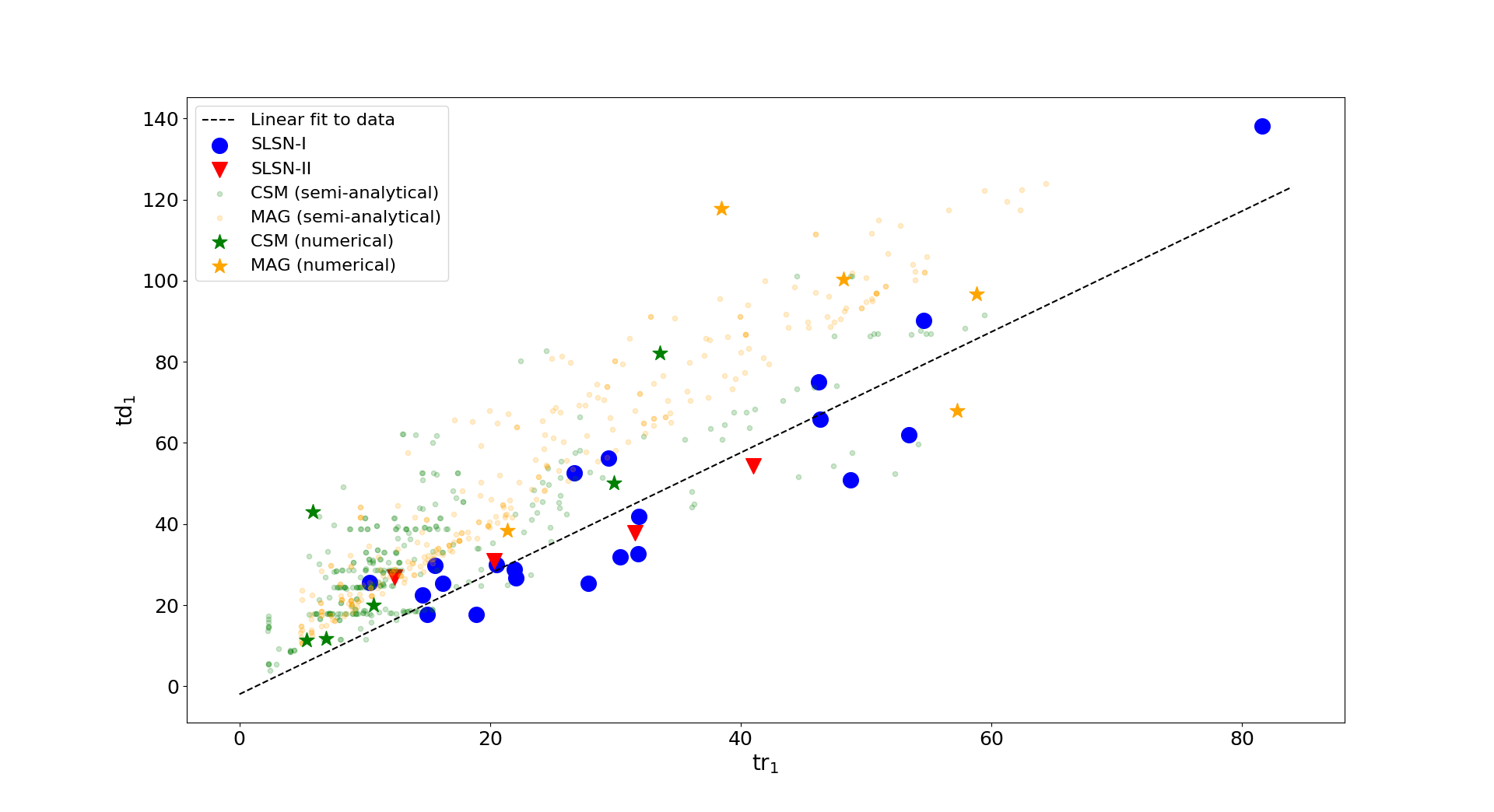}
\caption{Distribution of $tr_{\rm 1}$ and $td_{\rm 1}$ for the semi--analytical MAG (orange circles) and CSM (green circles) 
models compared to the observed SLSN--I (blue circles) and SLSN--II (red triangles) sample. The green and orange star symbols
correspond to published numerical LC models (see \ref{mod}). The dashed line represents a linear fit to the observed data.}
\label{Fig:tr1td1fit}
\end{center}
\end{figure}

\begin{figure}
\begin{center}
\includegraphics[angle=0,width=9cm]{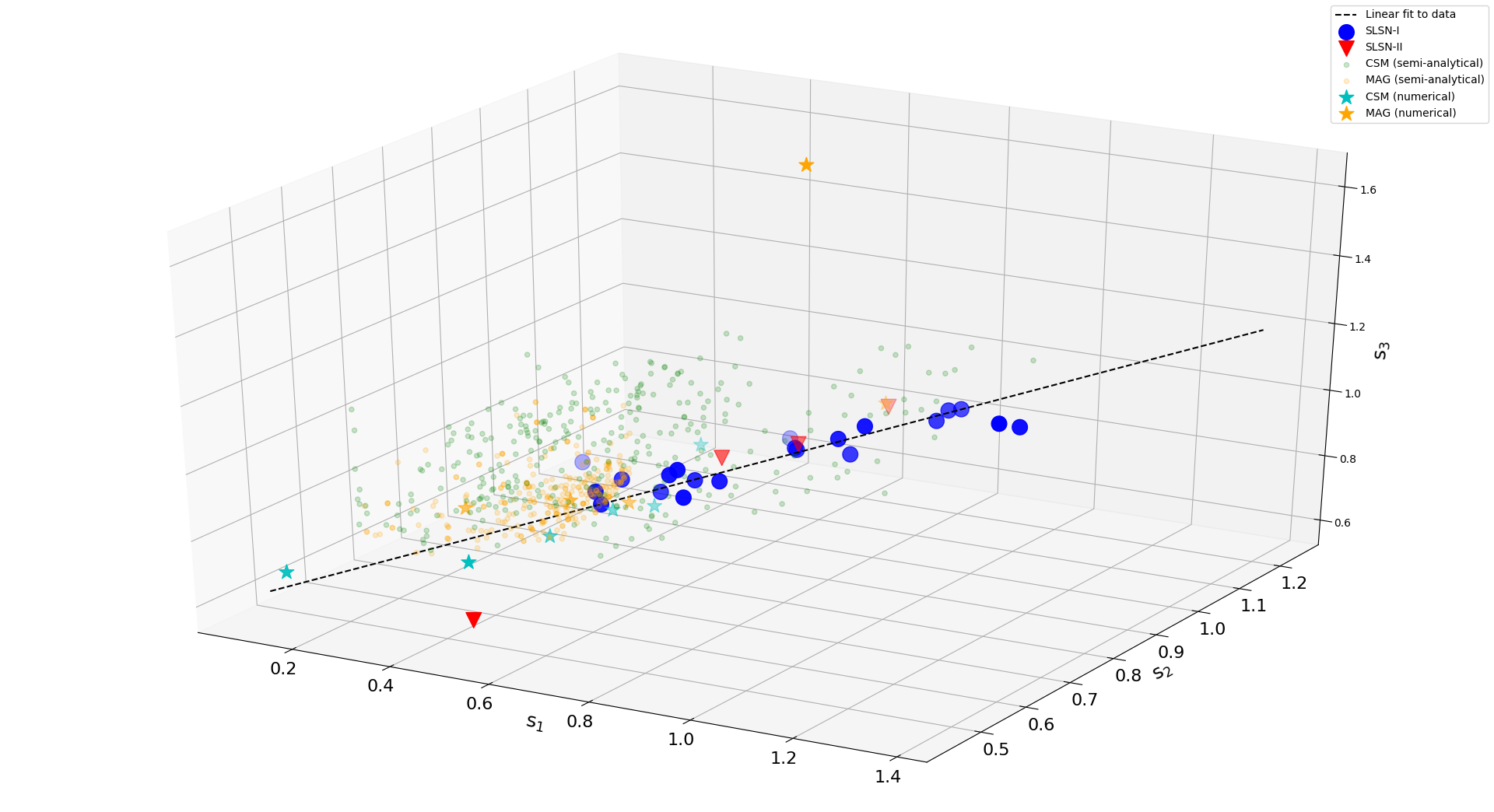}
\caption{Same as Figure~\ref{Fig:tr1td1fit} but for $s_{\rm 1}$, $s_{\rm 2}$ and $s_{\rm 3}$.}
\label{Fig:s1s2s3fit}
\end{center}
\end{figure}

For the CSM model we consider cases with H--poor opacity (CSM--I; $\kappa =$~0.2~cm$^{2}$~g$^{-1}$)
and H--rich opacity ($\kappa =$~0.4~cm$^{2}$~g$^{-1}$) and run two sets of grids:
(a) CSM--I$\kappa$/CSM--II$\kappa$ models, where the parameter grid is identical and
(b) CSM--I/CSM--II models where the parameter grid is constrained in each case,
motivated by assumptions about the nature of the progenitor stars in Type I versus Type II SNe respectively
that are further discussed later in this section.
For case (a) the ranges used for each parameter are as following:
\begin{itemize}
\item{$E_{\rm SN, 51}  \in [1.0,1.2,1.5,2.0]$, where $E_{\rm SN} = E_{\rm SN,51} \times 10^{51}$~erg}
\item{$M_{\rm ej}  \in [5,8,10,15,20,25,30,40]$, where $M_{\rm ej}$ is in units of $M_{\odot}$}
\item{$n \in [7,8,9,10,11,12]$}
\item{$R_{\rm CS,15}  \in [10^{-5},10^{-4},10^{-3},10^{-2},10^{-1}]$, where $R_{\rm CS} = R_{\rm CS,15} \times 10^{15}$~cm}
\item{$M_{\rm CS}  \in [0.1,0.2,0.5,1.0,2.0,5.0,8.0]$, where $M_{\rm CS}$ is in units of $M_{\odot}$}
\item{$\dot{M}  \in [0.001,0.01,0.05,0.1,0.2,0.5,1]$, where $\dot{M}$ is in units of $M_{\odot}$~yr$^{-1}$.}
\end{itemize}

For case (b) and the CSM--I subset, the ranges used are:
\begin{itemize}
\item{$E_{\rm SN, 51}  \in [1,1.2,1.5,1.75,2]$}
\item{$M_{\rm ej}  \in [5,8,10,12,15,20,25,30]$}
\item{$n \in [7,8,9]$}
\item{$R_{\rm CS,15}  \in [10^{-5},10^{-4},5 \times 10^{-4},10^{-3},5 \times 10^{-3},10^{-2}]$}
\item{$M_{\rm CS}  \in [0.1,0.2,0.5,0.7,1.0,2.0,5.0]$}
\item{$\dot{M}  \in [10^{-5},10^{-4},10^{-3},0.01,0.1,0.2,0.5,1.0,2.0]$,}
\end{itemize}
and accordingly for the CSM--II subset:
\begin{itemize}
\item{$E_{\rm SN, 51}  \in [1,1.2,1.5,1.75,2]$}
\item{$M_{\rm ej}  \in [12,15,20,25,30,40,50,60]$}
\item{$n \in [10,11,12]$}
\item{$R_{\rm CS,15}  \in [0.01,0.05,0.08,0.10,0.20,0.30]$}
\item{$M_{\rm CS}  \in [0.5,1.0,2.0,5.0,8.0,10.0,15.0]$}
\item{$\dot{M}  \in [10^{-5},10^{-4},10^{-3},0.01,0.1,0.2,0.5,1.0,2.0]$}
\end{itemize}
For all CSM models we are focusing on the $s =$~0 cases implying a fiducial, constant--density circumstellar shell. While
the $s =$~2 case is of interest since it implies a radiatively--driven wind structure that is common around
red supergiant stars (RSGs) we omit it in this work because it is inconsistent with episodic mass--loss, that is more
likely to be the case for luminous SNe. Also, for the vast majority of cases where the $s =$~2 choice yields luminous LCs
other parameters obtain unrealistic values (for example, $M_{\rm CS}$ values in excess of $\sim$~100~$M_{\odot}$ are
commonly found; C13). As a result, a total of 47,040 models were generated for the CSM--I$\kappa$/CSM--II$\kappa$ cases
and 45,360 models for the CSM--I/CSM--II cases. 

\setcounter{table}{2}
\begin{deluxetable*}{lccccc|ccccccc}
\tabletypesize{\footnotesize}
\tablecaption{Main statistical properties of the CSM--I and CSM --II model samples used in this work. \label{T3}}
\tablehead{
\colhead {Parameter} &
\colhead{$\mu$} &
\colhead{$M$} &
\colhead {$\sigma$} &
\colhead{max} &
\colhead {min} &
\colhead{$\mu$} &
\colhead{$M$} &
\colhead {$\sigma$} &
\colhead{max} &
\colhead {min} &
\\}
\startdata
&&& CSM--I & & & & & CSM--II & & \\
\hline
$tr_{\rm 1}$ & 12.2 & 11.0 & 5.9 & 36.1 & 2.3  & 45.1 & 46.6 & 8.8 & 59.5 & 17.3 \\
$td_{\rm 1}$ & 29.7 & 28.6 & 13.2 & 82.8 & 4.0 & 72.6 & 69.5 & 16.1 & 101.1 & 44.9 \\
$s_{\rm 1}$ & 0.43 & 0.41 & 0.15 & 0.87 & 0.13 &  0.64 & 0.61 & 0.13 & 1.00 & 0.37 \\
$tr_{\rm 2}$ & 7.0 & 6.0 & 4.0 & 28.2 & 1.3 &  18.7 & 19.4 & 3.7 & 25.4 & 7.1 \\
$td_{\rm 2}$& 9.1 & 8.4 & 5.0 & 33.3 & 1.5 & 24.4 & 25.9 & 4.7 & 31.7 & 11.8 \\
$s_{\rm 2}$ & 0.78 & 0.77 & 0.15 & 1.15 & 0.52 & 0.77 & 0.74 & 0.14 & 1.14 & 0.60 \\
$tr_{\rm 3}$ & 2.9 & 2.2 & 2.6 & 18.9 & 0.5 & 6.1 & 6.1 & 1.2 & 8.3 & 2.5 \\
$td_{\rm 3}$ & 3.1 & 2.4 & 3.1 & 24.4 & 0.5  & 7.0 & 7.2 & 1.4 & 10.2 & 3.0 \\
$s_{\rm 3}$ & 0.92 & 0.93 & 0.11 & 1.10 & 0.73 & 0.88 & 0.85 & 0.10 & 1.09 & 0.73 \\
\hline
\enddata 
\end{deluxetable*}

\setcounter{table}{3}
\begin{deluxetable*}{lccccc|ccccccc}
\tabletypesize{\footnotesize}
\tablecaption{Main statistical properties of the CSM--I$\kappa$ and CSM --II$\kappa$ model samples used in this work.\label{T4}}
\tablehead{
\colhead {Parameter} &
\colhead{$\mu$} &
\colhead{$M$} &
\colhead {$\sigma$} &
\colhead{max} &
\colhead {min} &
\colhead{$\mu$} &
\colhead{$M$} &
\colhead {$\sigma$} &
\colhead{max} &
\colhead {min} &
\\}
\startdata
&&& CSM--I$\kappa$ & & & & & CSM--II$\kappa$ & & \\
\hline
$tr_{\rm 1}$ & 15.1 & 12.6 & 8.3 & 50.3 & 2.5 & 11.9 & 11.5 & 3.5 & 22.1 & 3.3 \\
$td_{\rm 1}$ & 32.2 & 30.5 & 16.0 & 83.2 & 3.3 & 25.3 & 22.9 & 10.7 & 49.7 & 5.0 \\
$s_{\rm 1}$ & 0.50 & 0.48 & 0.18 & 1.03 & 0.16 & 0.52 & 0.48 & 0.16 & 0.86 & 0.26 \\
$tr_{\rm 2}$ & 7.7 & 6.1 & 5.4 & 39.3 & 1.7 & 6.8 & 6.3 & 3.5 & 20.8 & 2.2 \\
$td_{\rm 2}$ & 10.4 & 8.5 & 6.4 & 38.9 & 1.9 & 8.4 & 7.5 & 4.0 & 22.3 & 1.9 \\
$s_{\rm 2}$ & 0.75 & 0.72 & 0.26 & 1.16 & 0.53 & 0.82 & 0.80 & 0.16 & 1.15 & 0.55 \\
$tr_{\rm 3}$ & 3.1 & 2.3 & 3.4 & 26.3 & 0.6 & 2.6 & 2.1 & 2.4 & 15.2 & 0.9 \\
$td_{\rm 3}$ & 3.6 & 2.5 & 4.2 & 32.5 & 0.6 & 2.9 & 2.5 & 2.9 & 18.6 & 1.0 \\
$s_{\rm 3}$ & 0.90 & 0.89 & 0.10 & 1.10 & 0.74 & 0.90 & 0.89 & 0.10 & 1.09 & 0.74 \\
\hline
\enddata 
\end{deluxetable*}

\setcounter{table}{4}
\begin{deluxetable}{lcccccc}
\tabletypesize{\footnotesize}
\tablecaption{Main statistical properties of the MAG model samples used in this work.\label{T5}}
\tablehead{
\colhead {Parameter} &
\colhead{$\mu$} &
\colhead{$M$} &
\colhead {$\sigma$} &
\colhead{max} &
\colhead {min} &
\\}
\startdata
&&& MAG & & \\
\hline
$tr_{\rm 1}$ & 22.8 & 18.7 & 14.3 & 64.4 & 4.9 \\
$td_{\rm 1}$ & 50.8 & 43.3 & 28.4 & 123.9 & 10.7 \\
$s_{\rm 1}$ & 0.44 & 0.46 & 0.08 & 0.54 & 0.20 \\
$tr_{\rm 2}$ & 15.2 & 12.5 & 9.3 & 41.4 & 3.3 \\
$td_{\rm 2}$ & 22.2 & 18.4 & 13.0 & 56.4 & 4.7 \\
$s_{\rm 2}$ & 0.68 & 0.69 & 0.05 & 0.78 & 0.52 \\
$tr_{\rm 3}$ & 8.8 & 7.2 & 5.3 & 23.5 & 1.9 \\
$td_{\rm 3}$ & 10.5 & 8.7 & 6.4 & 27.1 & 2.06 \\
$s_{\rm 3}$ & 0.85 & 0.84 & 0.07 & 1.09 & 0.73 \\
\hline
\enddata
\end{deluxetable}

\setcounter{table}{5}
\begin{deluxetable*}{lccccccccccccc}
\tabletypesize{\footnotesize}
\tablecaption{The numerical models sample.\label{T6}}
\tablehead{
\colhead {Model ID} &
\colhead{Reference} &
\colhead{Model Type} &
\colhead {$tr_{\rm 1}$} &
\colhead {$td_{\rm 1}$} &
\colhead {$s_{\rm 1}$} &
\colhead {$tr_{\rm 2}$} &
\colhead {$td_{\rm 2}$} &
\colhead {$s_{\rm 2}$} &
\colhead {$tr_{\rm 3}$} &
\colhead {$td_{\rm 3}$} &
\colhead {$s_{\rm 3}$} &
\\}
\startdata
\hline
{\tt B3} & \citet{2016MNRAS.tmp..117D} & CSM--I & 5.9 & 43.0 & 0.14 & 4.3 & 9.8 & 0.44 & 2.7 & 3.9 & 0.70 \\
{\tt T130D-b} & \citet{2017ApJ...836..244W} & CSM--I & 6.9 & 11.9 & 0.59 & 4.3 & 5.8 & 0.75 & 2.3 & 2.9 & 0.80 \\
{\tt D2} & \citet{2013MNRAS.428.1020M} & CSM--II & 29.9 & 50.1 & 0.60 & 19.0 & 22.7 & 0.84 & 10.5 & 11.3 & 0.93 \\
{\tt F1} & \citet{2013MNRAS.428.1020M} & CSM--II & 33.5 & 82.0 & 0.41 & 23.3 & 43.1 & 0.54 & 13.7 & 18.8 & 0.73 \\
{\tt R3} & \citet{2016MNRAS.tmp..117D} & CSM--II & 5.4 & 11.4 & 0.47 & 3.7 & 5.7 & 0.65 & 2.0 & 2.7 & 0.75 \\
{\tt T20} & \citet{2017ApJ...836..244W} & CSM--II & 10.7 & 20.0 & 0.53 & 7.0 & 9.8 & 0.71 & 3.7 & 4.7 & 0.80 \\
\hline
{\tt KB 1} (Black curve) & \citet{2010ApJ...717..245K} & MAG & 21.4 & 38.5 & 0.56 & 13.7 & 18.8 & 0.73 & 7.7 & 9.4 & 0.82 \\
{\tt KB 2} (Red curve) & \citet{2010ApJ...717..245K} & MAG & 38.5 & 117.9 & 0.33 & 25.37 & 40.9 & 0.62 & 14.7 & 18.0 & 0.82 \\
{\tt Model 2} & \citet{2016ApJ...821...36K} & MAG & 48.2 & 100.3 & 0.48 & 33.6 & 49.3 & 0.68 & 20.2 & 24.1 & 0.84 \\
{\tt RE3B1} & \citet{dessartaudit} & MAG & 58.8 & 96.7 & 0.61 & 46.5 & 43.9 & 1.06 & 31.1 & 19.2 & 1.62 \\
{\tt RE0p4B3p5} & \citet{dessartaudit} & MAG & 57.3 & 68.0 & 0.84 & 34.8 & 35.6 & 0.97 & 19.0 & 18.6 & 1.02 \\
\enddata
\end{deluxetable*}

\setcounter{table}{6}
\begin{deluxetable*}{lccccc|ccccccc}
\tabletypesize{\footnotesize}
\tablecaption{Main statistical properties of the numerical models.\label{T7}}
\tablehead{
\colhead {Parameter} &
\colhead{$\mu$} &
\colhead{$M$} &
\colhead {$\sigma$} &
\colhead{max} &
\colhead {min} &
\colhead{$\mu$} &
\colhead{$M$} &
\colhead {$\sigma$} &
\colhead{max} &
\colhead {min} &
\\}
\startdata
&&& CSM--I/CSM--II & & & & & MAG & & \\
\hline
$tr_{\rm 1}$ & 15.4 & 8.8 & 33.5 & 5.4 & 11.7 & 44.8 & 48.2 & 58.8 & 22.4 & 13.8 \\
$td_{\rm 1}$ & 36.4 & 31.5 & 82.1 & 11.4 & 25.2 & 84.3 & 96.7 & 117.9 & 38.5 & 28.0 \\
$s_{\rm 1}$ & 0.46 & 0.50 & 0.60 & 0.14 & 0.16 & 0.56 & 0.56 & 0.84 & 0.33 & 0.17 \\
$tr_{\rm 2}$ & 10.3 & 5.7 & 23.3 & 3.7 & 7.9 & 30.8 & 33.6 & 46.5 & 13.7 & 10.9 \\
$td_{\rm 2}$ & 16.13 & 9.8 & 43.1 & 5.7 & 13.3 & 37.7 & 40.9 & 49.3 & 18.8 & 10.5 \\
$s_{\rm 2}$ & 0.66 & 0.68 & 0.84 & 0.44 & 0.133 & 0.81 & 0.73 & 1.06 & 0.62 & 0.17 \\
$tr_{\rm 3}$ & 5.8 & 3.2 & 13.7 & 2.7 & 4.6 & 18.5 & 19.0 & 31.1 & 7.7 & 7.7 \\
$td_{\rm 3}$ & 7.4 & 4.3 & 18.8 & 2.7 & 5.9 & 17.9 & 18.6 & 24.1 & 9.4 & 4.8 \\
$s_{\rm 3}$ & 0.79 & 0.78 & 0.93 & 0.70 & 0.07 & 1.02 & 0.84 & 1.62 & 0.82 & 0.31 \\
\hline
\enddata 
\end{deluxetable*}

Our motivation for adopting different parameter ranges for the CSM--I and CSM--II models stems from several factors.
First, larger $M_{\rm CS}$ values are possible in the CSM--II case as suggested by spectroscopic observations of SLSN-II of Type IIn
\citep{2010ApJ...709..856S} where stronger mass--loss pertains due to LBV--type or PPISN processes \citep{2014ARA&A..52..487S}.
That, in turn, also implies larger progenitor masses (and therefore $M_{\rm ej}$) for CSM--II, 
as is the case for regular luminosity SNe where LC fits imply larger $M_{\rm ej}$, and therefore larger diffusion timescales, for Type II events
than for Type I SNe. Finally, lower values of $n$ are more typical of compact, blue supergiant (BSG) progenitors with radiative envelopes while
while higher values imply extended, RSG--type convective envelopes that are more appropriate for SLSN--II \citep{2003LNP...598..171C}.
In summary, the CSM--II parameters are associated with RSG--type progenitors with extended H--rich envelopes while the CSM--I parameters
with more compact, BSG--type stars.

We caution that one potential issue with our choices for model parameter grids, is there are no good observational constrains yet 
on what the shape of the distribution of SN ejecta and circumstellar shell masses should be, so using these models in a clustering analysis (Section~\ref{cluster}) 
might be misleading as it can create dense clusters of models that might actually be very sparsely populated in nature, 
or conversely an underdensity of points in regions where more MAG or CSM SNe might lie in reality. 
Our grid selection for $M_{\rm CS}$ is largely driven by published observations of nebular shells around massive, LBV--type stars indicating $M_{\rm CS} \simeq$~0.1--20~$M_{\odot}$ 
\citep{2006ApJ...645L..45S,2009ApJ...705L..25G,2010AJ....139.2330W,2010MNRAS.403..760G,2014ARA&A..52..487S}. The ranges for
$M_{\rm ej}$ are within typical ranges for stars massive enough to experience a SN, and in agreement with observations of SN progenitor
stars in pre--explosion images and supernova remnants ($M_{\rm ej} \simeq$~8--25~$M_{\odot}$) 
\citep{2009ARA&A..47...63S,2018ApJ...858...15M,2019ApJ...871...64A}. Higher--mass progenitors cannot be excluded given observations
of stars as massive as $>$~150~$M_{\odot}$ in the Milky Way galaxy \citep{2010MNRAS.408..731C}.

For the MAG model, we investigate a dense grid of models with $10^{12} < B_{\rm MAG} < 10^{15}$~G and $1.0 < P_{\rm MAG} < 50$~ms, where
$B_{\rm MAG}$ and $P_{\rm MAG}$ is the magnetic field and the initial rotational period of the magnetar respectively. We are also varying the
diffusion timescale, $t_{\rm d}$, that further controls the shape of MAG model LCs (Equation 13 of C12), in the range $3 < t_{\rm d} < 100$~days.
The grid resolution we use for these parameters results to a total of 46,656 MAG model LCs generated.

A large fraction of CSM and MAG models did not produce superluminous LCs, which we take to be those reaching $L_{\rm max} = 10^{44}$~erg~s$^{-1}$ 
or more \citep{2012Sci...337..927G}. These models are ignored from each of our CSM and MAG model samples for further analysis. In addition, we exclude model LCs
that result in physically inconsistent parameters such as combinations of $B_{\rm MAG}$ and $P_{\rm MAG}$ values in the MAG model that 
are incompatible with the convective dynamo process in magnetars \citep{1992ApJ...392L...9D}, and CSM models that yield $M_{\rm CS}$ too large compared
to the associated $M_{\rm ej}$ values that represent a measure of the total progenitor mass. 

As a result, our original CSM--I/CSM--II, CSM--I$\kappa$/CSM--II$\kappa$ and MAG model samples are each reduced into smaller subsamples of nearly equal size
that are then used in our final LC shape parameter analysis. 
More specifically, a total of 306 CSM--I/CSM--II, 248 CSM--I$\kappa$/CSM--II$\kappa$ and 304 MAG superluminous LC models are used in this work. The
statistical properties of the LC shape parameters of all models are summarized in Tables~\ref{T3} through~\ref{T5}. Figures~\ref{Fig:tr1td1} and~\ref{Fig:s1s2s3}
show the distribution of a few LC shape parameters ($tr_{\rm 1}$, $td_{\rm 1}$, $s_{\rm 1}$, $s_{\rm 2}$, $s_{\rm 3}$) for the CSM--I/CSM--II and MAG model 
samples and Figure~\ref{Fig:symmLCs} examples of some of the most symmetric LCs in these samples.

For comparison against our semi--analytical LCs, we have also included a sample numerical CSM and MAG LCs available in the literature. Table~\ref{T6} lists the details
of the numerical model LCs and Table~\ref{T7} summarizes the statistics of their shape parameters. 
Figure~\ref{Fig:tr1td1fit} is a scatter plot between $tr_{\rm 1}$ and $td_{\rm 1}$ for all samples in this work, including the numerical MAG and CSM models. A linear best--fit
to the observed SLSN--I and SLSN--II data is also shown (see Equation~\ref{Eq1}). Although we chose to not use different symbols for the CSM models as
presented in Figure~\ref{Fig:tr1td1fit}, it is evident by inspecting Table~\ref{T4} that CSM--II models occupy the upper right corner of this plot given their
longer primary rise and decline timescales. A few SLSN--I thus appear to be associated with the CSM--II data that were chosen based on assumptions for the
progenitors of H--rich SLSNe. The situation is different when looking at the CSM--I$\kappa$/CSM--II$\kappa$ distribution, however, where the parameter
grids are identical and the only difference is due to different SN ejecta + CS shell opacity. In this case, the primary timescales of the models are consistent.
Very slowly evolving H--poor SLSNe may be hard to produce under the assumption of H--poor CSM interaction given the large, 
H--deficient CS shell mass needed to account for the long primary rise and decline timescales. Interaction with a H--poor CS shells of non--spherical geometry in combination
with viewing--angle effects may be a way out of this apparent discrepancy \citep{2018MNRAS.475.3152K}.
Accordingly, Figure~\ref{Fig:s1s2s3fit} shows a 3D scatter plot for the primary, secondary and tertiary LC symmetry parameter for all samples.
The superluminous LCs recovered infer the following mean values for the parameters of each model:
\begin{itemize}
\item{CSM--I: $E_{\rm SN, 51} =$~1.75, $M_{\rm ej} =$~10~$M_{\odot}$, $n =$~8, $R_{\rm CS,15} =$~0.006, $M_{\rm CS} =$~1~$M_{\odot}$ and $\dot{M} =$~0.01~$M_{\odot}$~yr$^{-1}$,}
\item{CSM--II: $E_{\rm SN, 51} =$~2.00, $M_{\rm ej} =$~13~$M_{\odot}$, $n =$~12, $R_{\rm CS,15} =$~0.2, $M_{\rm CS} =$~10~$M_{\odot}$ and $\dot{M} =$~0.01~$M_{\odot}$~yr$^{-1}$,}
\item{CSM--I$\kappa$: $E_{\rm SN, 51} =$~1.80, $M_{\rm ej} =$~10~$M_{\odot}$, $n =$~9, $R_{\rm CS,15} =$~0.08, $M_{\rm CS} =$~2~$M_{\odot}$ and $\dot{M} =$~0.15~$M_{\odot}$~yr$^{-1}$,}
\item{CSM--II$\kappa$: $E_{\rm SN, 51} =$~2.00, $M_{\rm ej} =$~7~$M_{\odot}$, $n =$~9, $R_{\rm CS,15} =$~0.1, $M_{\rm CS} =$~0.3~$M_{\odot}$ and $\dot{M} =$~0.3~$M_{\odot}$~yr$^{-1}$,}
\item{MAG:  $B_{\rm MAG} = 1.4 \times 10^{13}$~G and $P_{\rm MAG} =$~1.3~ms.}
\end{itemize}
These parameters are within the range of semi--analytical and numerical fits of the CSM and MAG models to observed SLSN LCs commonly found in the literature.

A careful examination of the computed LC shape parameter distributions for the CSM and MAG models reveals a lot of interesting insights. First, the primary
rise and decline timescales appear to have a binary distribution for the CSM models with CSM--I models typically reaching shorter $tr_{\rm 1}$ and $td_{\rm 1}$ values
than CSM--II models. This is both due to the physically--motivated choices for the parameter grids discussed earlier, but also because of the opacity difference between
H--rich and H--poor models. On the other hand, the MAG models show a more continuous and single--peaked distribution with typical values 
$tr_{\rm 1} \simeq$~5--15~days and $td_{\rm 1} \simeq$~20--30~days. In terms of LC symmetry, the majority of models do not appear to produce symmetric
LCs around the primary luminosity threshold as $0.9 < s_{\rm 1} < 1.1$ values are rarely recovered. In fact, CSM is the only set of models reaching $s_{\rm 1}$ values
close to unity while MAG is unable to produce any models with symmetric LCs both in terms of $s_{\rm 1}$ and $s_{\rm 2}$. Even the most
symmetric MAG LCs in our sample appear to have this issue (Figure~\ref{Fig:symmLCs})
{\it This is an important issue for MAG models given that a significant fraction of observed SLSN--I are symmetric around these luminosity thresholds} (Section~\ref{obs}).
This seems to be the case for numerically--computed MAG LC models as well, with the most symmetric one being model {\tt RE0p4B3p5} 
\citep{dessartaudit} with $s_{\rm 1} =$~0.84. Numerical CSM models tend to yield more rapidly--evolving LCs than their semi--analytical counterparts. The primary
source of this difference is the assumption of a constant diffusion timescale in the semi--analytical CSM models \citep{2013MNRAS.428.1020M,2018arXiv181206522K}.

We explore the possiblity that gamma--ray leakage produces faster--declining MAG LCs, therefore enhancing symmetry, by adopting the same
formalism employed in the case of LCs powered by the radioactive decay of $^{56}$Ni \citep{1984ApJ...280..282S,1997ApJ...491..375C,2008MNRAS.383.1485V,2013ApJ...773...76C}.
Using a fiducial SN ejecta gamma--ray opacity of $\kappa_{\rm \gamma} =$~0.03~cm$^{2}$~g$^{-1}$ and the implied SN ejecta mass for the two most symmetric MAG models
shown in the top right panel of Figure~\ref{Fig:symmLCs}, we adjust the output luminosity as $L^{\prime}(t) = L(t) (1-\exp{-A t^{-2}})$, where $A t^{-2} = \kappa_{\rm \gamma} \rho R$.
The two most symmetric MAG models with high gamma--ray leakage are then plotted as dashed curves. Allowing for gamma--rays to escape can increase the
decline rate of the LC at late times leading to shorter $td_{\rm 1}$ and slightly higher $s_{\rm 1}$ values. The change, however, still falls short in producing symmetric MAG
LCs since $s_{\rm 1}$ only increases by 14--22\% and the maximum value for $s_{\rm 1} \lessapprox$~0.6.

Second, the observed tight $tr_{\rm 1}$--$td_{\rm 1}$ correlation in SLSN LCs is reproduced by both CSM and MAG models. CSM models generally 
predict faster--evolving LCs at late times than MAG models, consistent with the observations. This is mainly due 
to the continuous power input in the MAG model that sustains a flatter LC at late times while in the CSM model the energy input is terminated abruptly leading
to rapid decline after peak luminosity (C12). An example of a SLSN with a very flat late--time LC is SN2015bn \citep{2018ApJ...866L..24N}, indicating
that this may be a good candidate for the MAG model.
The observed LC symmetry parameter distributions (Figure~\ref{Fig:s1s2s3fit}) reveal a more distinct dichotomy between CSM and MAG models. MAG models 
fail to produce fully symmetric LCs and are clustered in a confined region of the 3D ($s_{\rm 1}$, $s_{\rm 2}$ and $s_{\rm 3}$) parameter space while CSM models
more scatter. 

Finally, we estimate the fraction of CSM and MAG model SLSN LCs that have a concave--up shape during the rise to peak luminosity or, in other words, positive
second derivative for $t<t_{\rm max}$. An example of an observed SLSN with concave--up LC during the rise is SN~2017egm \citep{2017ApJ...851L..14W}.
Not a single MAG LCs is found to be concave--up during the rise. On the contrary, $\sim$~20\% of CSM--I, $\sim$~60\% of CSM--II and
$\sim$~50\% of CSM--I$\kappa$/CSM--II$\kappa$ models are found to have concave--up rise to peak luminosity. The implication is that
the shape of the rising part of SLSN LCs may also be tied to the nature of the power input mechanism and, specifically, the functional form of the input luminosity.
Continuous, monotonically declining power inputs like $^{56}$Ni decay and magnetar spin--down energy correspond to concave--down SLSN LCs while
truncated CSM shock luminosity input depends on the details of the SN ejecta and the circumstellar material density structure and can yield either concave--up
or concave--down LCs during the early, rising phase.
This further enforces the need to obtain high--cadence photometric coverage of these events in the future transient surveys. 

\setcounter{table}{7}
\begin{deluxetable*}{lccccccccc}
\tabletypesize{\footnotesize}
\tablecaption{Details of clustering analysis.}
\tablehead{
\colhead{Datasets} &
\colhead {Parameters} &
\colhead{$N_{\rm D}$$^{\ast}$} &
\colhead{$k$} &
\colhead {$E_{\rm N}$} &
\colhead{$\bar{S}$} &
\colhead {$C_{\rm 0}$$^{\dagger}$} &
\colhead {$C_{\rm 1}$$^{\dagger}$} &
\colhead {$C_{\rm 2}$$^{\dagger}$} &
\\}
\startdata
\hline
CSM--I/CSM--II/MAG & $tr_{\rm 1}$,$td_{\rm 1}$ & 2 & 2 & 0.62 & 0.66 & 5.95/18.45/75.6 & 59.28/0.68/40.05 & - \\
& & & & & & 33.33/25.00 & 66.67/75.00 & - \\
CSM--I/CSM--II/MAG & $tr_{\rm 1}$,$td_{\rm 1}$ & 2 & 3 & 0.46 & 0.58 & 61.11/0.00/38.89 & 27.70/12.16/60.14 & 0.00/19.05/80.95 \\
& & & & & & 57.14/50.00 & 28.57/50.00 & 14.29/0.00 \\
CSM--I/CSM--II & $tr_{\rm 1}$,$td_{\rm 1}$ & 2 & 2 & 0.77 & 0.63 & 99.59/0.41 & 48.44/51.56 & - \\
& & & & & & 57.14/75.00 & 42.86/25.00 & - \\
CSM--I$\kappa$/CSM--II$\kappa$/MAG & $tr_{\rm 1}$,$td_{\rm 1}$ & 2 & 2 & 0.68 & 0.65 & 44.61/11.03/44.36 & 11.19/0.00/88.81 & - \\
& & & & & & 66.67/75.00 & 33.33/25.00 & - \\
CSM--I$\kappa$/CSM--II$\kappa$/MAG & $tr_{\rm 1}$,$td_{\rm 1}$ & 2 & 3 & 0.49 & 0.56 & 38.89/1.85/59.26 & 42.90/13.53/43.56 & 1.30/0.0/98.70 \\
& & & & & & 28.57/50.0 & 57.14/50.00 & 14.29/0.00 \\
CSM--I$\kappa$/CSM--II$\kappa$ & $tr_{\rm 1}$,$td_{\rm 1}$ & 2 & 2 & 0.66 & 0.57 & 77.18/22.82 & 88.76/11.24 & - \\
& & & & & & 47.62/50.00 & 52.38/50.00 & - \\
\hline
CSM--I/CSM--II/MAG & $s_{\rm 1}$,$s_{\rm 2}$,$s_{\rm 3}$ & 3 & 2 & $<$0.01 & 0.43 & 34.55/4.07/61.38 & 86.44/11.86/1.69 & - \\
& & & & & & 23.81/25.00 & 76.19/75.00 & - \\
CSM--I/CSM--II/MAG & $s_{\rm 1}$,$s_{\rm 2}$,$s_{\rm 3}$ & 3 & 3 & $<$0.01 & 0.32 & 26.19/4.76/69.05 & 82.35/17.65/0.00 & 71.34/2.44/26.22 \\
& & & & & & 28.57/25.00 & 71.43/75.00 & 0.00/0.00 \\
CSM--I/CSM--II & $s_{\rm 1}$,$s_{\rm 2}$,$s_{\rm 3}$ & 3 & 2 & $<$0.01 & 0.33 & 82.31/17.69 & 93.75/6.25 & - \\
& & & & & & 80.95/75.00 & 19.05/25.00 & - \\
CSM--I$\kappa$/CSM--II$\kappa$/MAG & $s_{\rm 1}$,$s_{\rm 2}$,$s_{\rm 3}$ & 3 & 2 & $<$0.01 & 0.60 & 31.12/5.81/63.07 & 73.33/26.67/0.00 & - \\
& & & & & & 42.86/25.00 & 57.14/75.00 & - \\
CSM--I$\kappa$/CSM--II$\kappa$/MAG & $s_{\rm 1}$,$s_{\rm 2}$,$s_{\rm 3}$ & 3 & 3 & $<$0.01 & 0.33 & 42.31/7.69/50.00 & 24.67/5.26/70.07 & 75.00/25.00/0.00 \\
& & & & & & 47.62/25.0 & 52.38/75.00 & 0.00/0.00 \\
CSM--I$\kappa$/CSM--II$\kappa$ & $s_{\rm 1}$,$s_{\rm 2}$,$s_{\rm 3}$ & 3 & 2 & $<$0.01 & 0.50 & 84.18/15.82 & 73.77/26.23 & - \\
& & & & & & 38.10/25.00 & 61.90/75.00 & - \\
\hline
CSM--I/CSM--II/MAG & $tr_{\rm 1}$,$td_{\rm 1}$,$tr_{\rm 2}$,$td_{\rm 2}$ & 4 & 2 & 0.71 & 0.66 & 2.44/18.90/78.66 & 60.09/0.67/39.24 & - \\
& & & & & & 38.10/25.00 & 61.90/75.00 & - \\
CSM--I/CSM--II/MAG & $tr_{\rm 1}$,$td_{\rm 1}$,$tr_{\rm 2}$,$td_{\rm 2}$ & 4 & 3 & 0.54 & 0.56 & 0.00/16.47/83.53 & 61.97/0.00/38.03 & 26.17/13.42/60.41 \\
& & & & & & 19.05/0.00 &52.38/50.00 & 28.57/50.00 \\
CSM--I/CSM--II & $tr_{\rm 1}$,$td_{\rm 1}$,$tr_{\rm 2}$,$td_{\rm 2}$ & 4 & 2 & 0.84 & 0.63 & 46.77/53.23 & 99.59/0.41 & - \\
& & & & & & 42.86/50.00 & 57.14/50.00 & - \\
CSM--I$\kappa$/CSM--II$\kappa$/MAG & $tr_{\rm 1}$,$td_{\rm 1}$,$tr_{\rm 2}$,$td_{\rm 2}$ & 4 & 2 & 0.77 & 0.64 & 44.81/11.14/44.05 & 11.56/0.00/88.44 & - \\
& & & & & & 61.90/75.00 & 38.10/25.00 & - \\
CSM--I$\kappa$/CSM--II$\kappa$/MAG & $tr_{\rm 1}$,$td_{\rm 1}$,$tr_{\rm 2}$,$td_{\rm 2}$ & 4 & 3 & 0.57 & 0.54 & 38.18/2.42/59.39 & 43.88/13.61/42.52 & 2.41/0.00/97.59 \\
& & & & & & 33.33/50.0 & 47.62/50.00 & 19.05/0.00 \\
CSM--I$\kappa$/CSM--II$\kappa$ & $tr_{\rm 1}$,$td_{\rm 1}$,$tr_{\rm 2}$,$td_{\rm 2}$ & 4 & 2 & 0.76 & 0.55 & 88.51/11.49 & 77.48/22.52 & - \\
& & & & & & 61.90/75.00 & 38.10/25.00 & - \\
\hline
CSM--I/CSM--II/MAG & $tr_{\rm 1}$,$td_{\rm 1}$,$tr_{\rm 2}$,$td_{\rm 2}$,$tr_{\rm 3}$,$td_{\rm 3}$ & 6 & 2 & 0.74 & 0.65 & 60.00/0.67/39.33 & 3.03/18.79/78.18 & - \\
& & & & & & 61.90/75.00 & 38.10/25.00 & - \\
CSM--I/CSM--II/MAG & $tr_{\rm 1}$,$td_{\rm 1}$,$tr_{\rm 2}$,$td_{\rm 2}$,$tr_{\rm 3}$,$td_{\rm 3}$ & 6 & 3 & 0.57 & 0.55 & 62.11/0.26/37.63 & 0.00/15.48/84.52 & 24.66/13.70/61.64 \\
& & & & & & 52.38/50.00 & 19.05/0.00 & 28.57/50.00 \\
CSM--I/CSM--II & $tr_{\rm 1}$,$td_{\rm 1}$,$tr_{\rm 2}$,$td_{\rm 2}$,$tr_{\rm 3}$,$td_{\rm 3}$ & 6 & 2 & 0.86 & 0.62 & 46.77/53.23 & 99.59/0.41 & - \\
& & & & & & 42.86/50.00 & 57.14/50.00 & - \\
CSM--I$\kappa$/CSM--II$\kappa$/MAG & $tr_{\rm 1}$,$td_{\rm 1}$,$tr_{\rm 2}$,$td_{\rm 2}$,$tr_{\rm 3}$,$td_{\rm 3}$ & 6 & 2 & 0.80 & 0.64 & 45.11/11.03/43.86 & 9.79/0.00/90.21 & - \\
& & & & & & 61.90/75.00 & 38.10/25.00 & - \\
CSM--I$\kappa$/CSM--II$\kappa$/MAG & $tr_{\rm 1}$,$td_{\rm 1}$,$tr_{\rm 2}$,$td_{\rm 2}$,$tr_{\rm 3}$,$td_{\rm 3}$ & 6 & 3 & 0.60 & 0.52 & 37.65/2.35/60.00 & 2.38/0.00/97.62 & 44.44/13.89/41.67 \\
& & & & & & 28.57/50.00 & 23.81/0.00 & 47.62/50.00 \\
CSM--I$\kappa$/CSM--II$\kappa$ & $tr_{\rm 1}$,$td_{\rm 1}$,$tr_{\rm 2}$,$td_{\rm 2}$,$tr_{\rm 3}$,$td_{\rm 3}$ & 6 & 2 & 0.82 & 0.54 & 77.18/22.82 & 88.76/11.24 & - \\
& & & & & & 33.33/25.00 & 66.67/75.00 & - \\
\hline
\enddata
\tablecomments{$^{\ast}$Normalized error ($E_{\rm N}$) values have been rounded to two decimal points. 
$^{\dagger}$The variables $C_{\rm 0}$, $C_{\rm 1}$, $C_{\rm 2}$ correspond to cluster associations with
SLSN LC models and the observed SLSN sample. The top line corresponds to percentages of model data, in the same order
as shown in the ``Models" column, that are assigned to the cluster. The bottom line corresponds to the percentage
of SLSN--I and SLSN--II (in the ``\%~SLSN--I/\%~SLSN--II" format) that are assigned to the cluster. 
\label{T8}}
\end{deluxetable*}

\begin{figure*}
\gridline{\fig{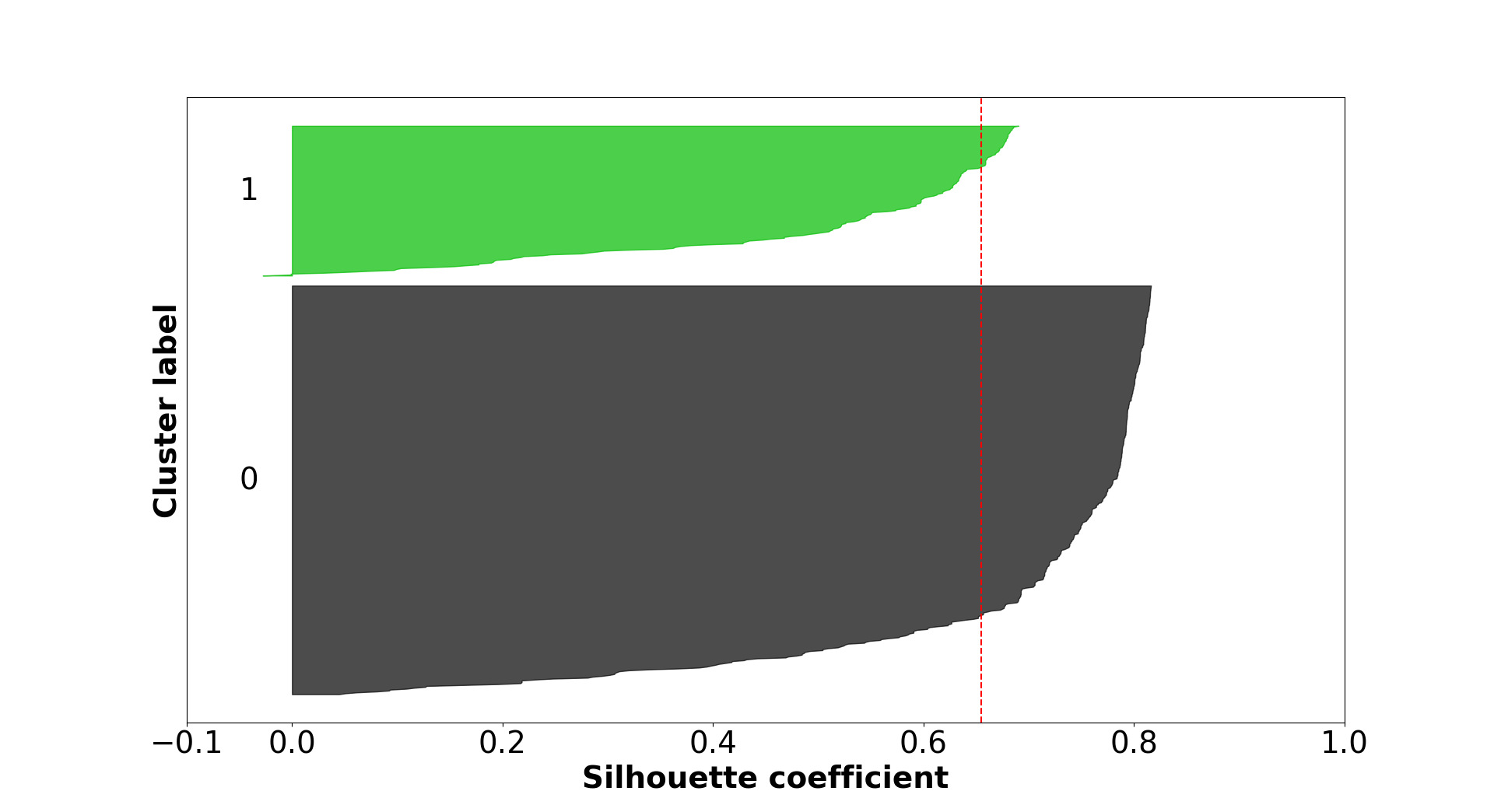}{0.5\textwidth}{}
          \fig{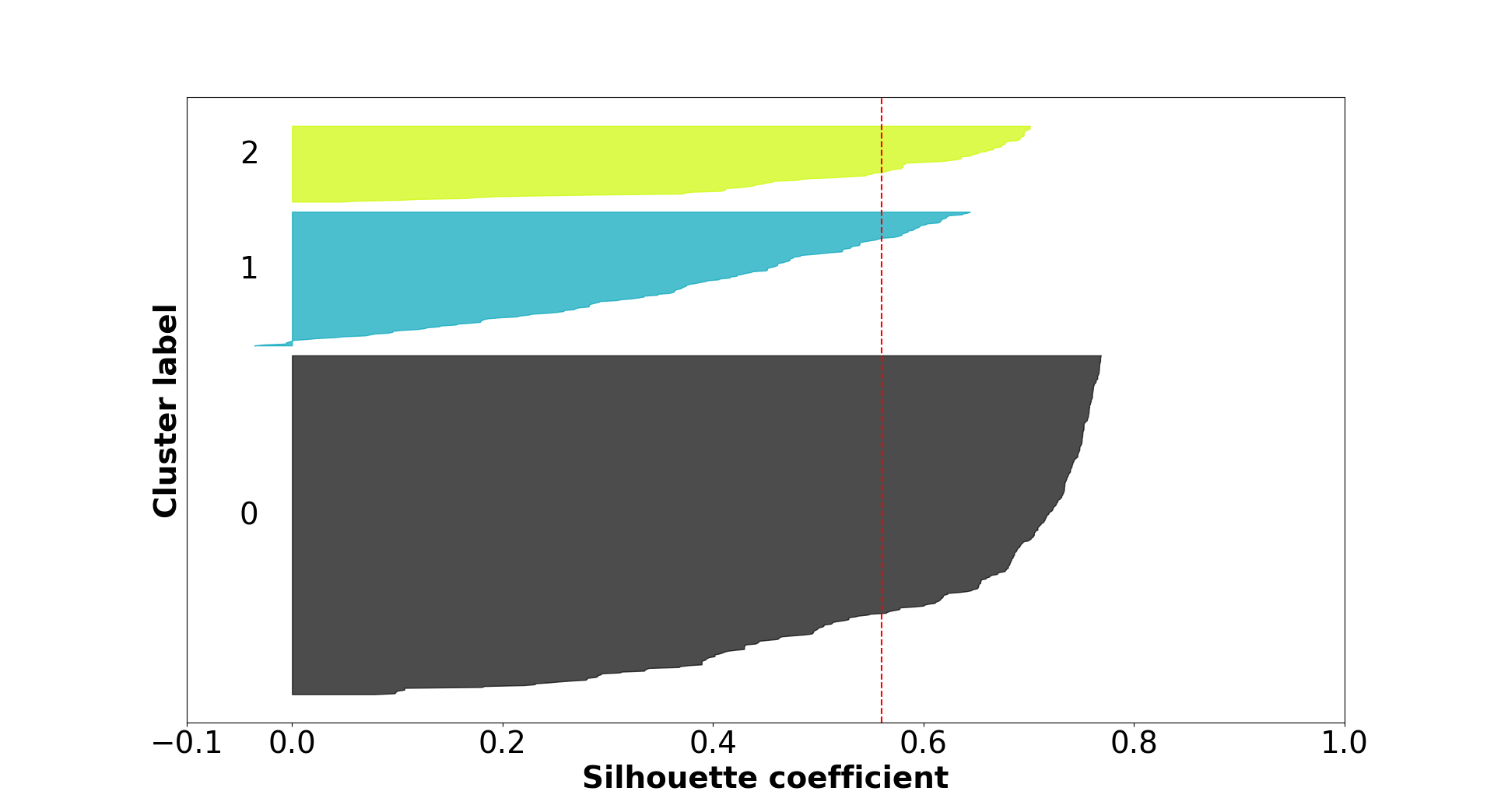}{0.5\textwidth}{}
          }
\gridline{\fig{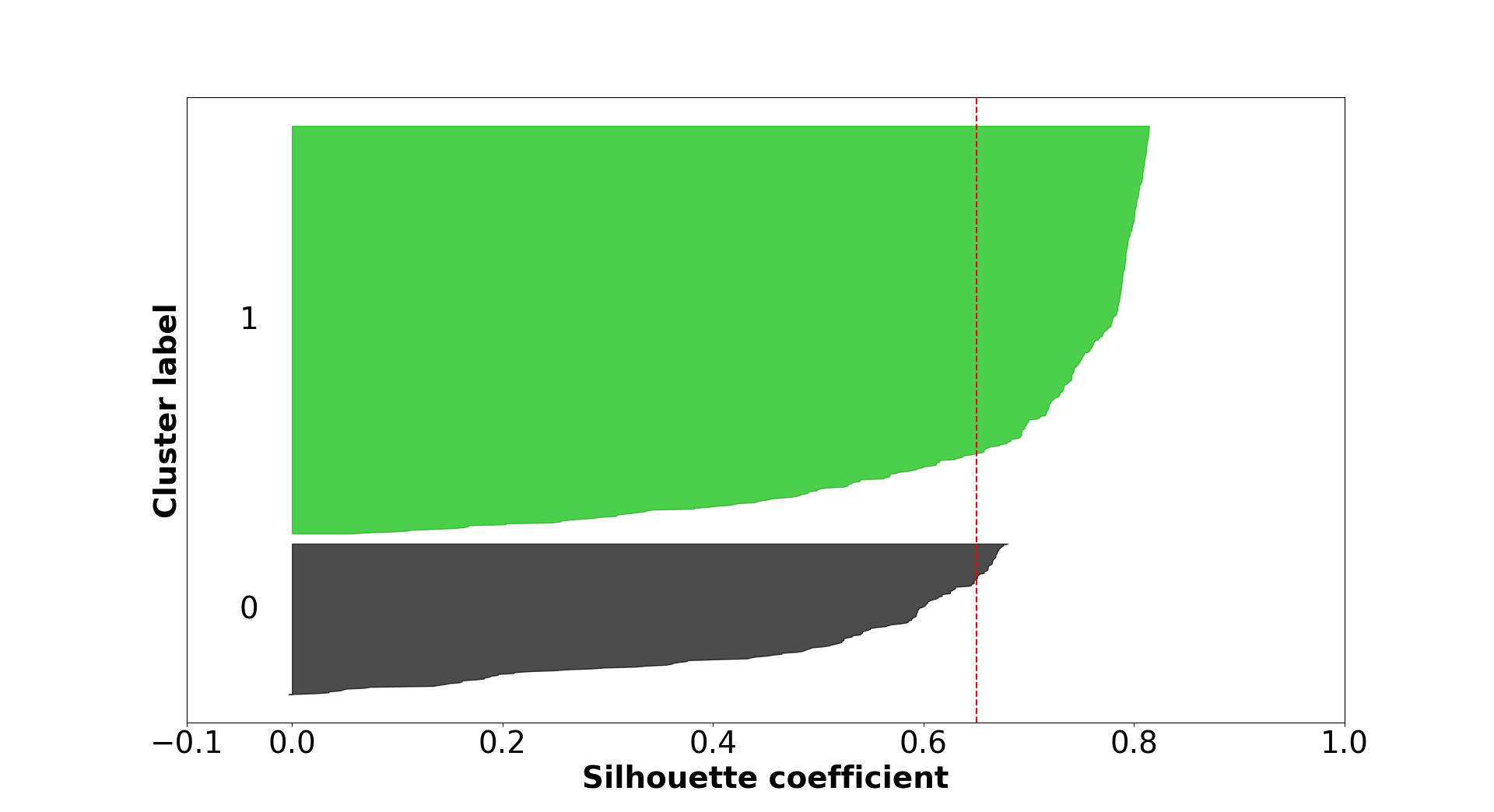}{0.5\textwidth}{}
         \fig{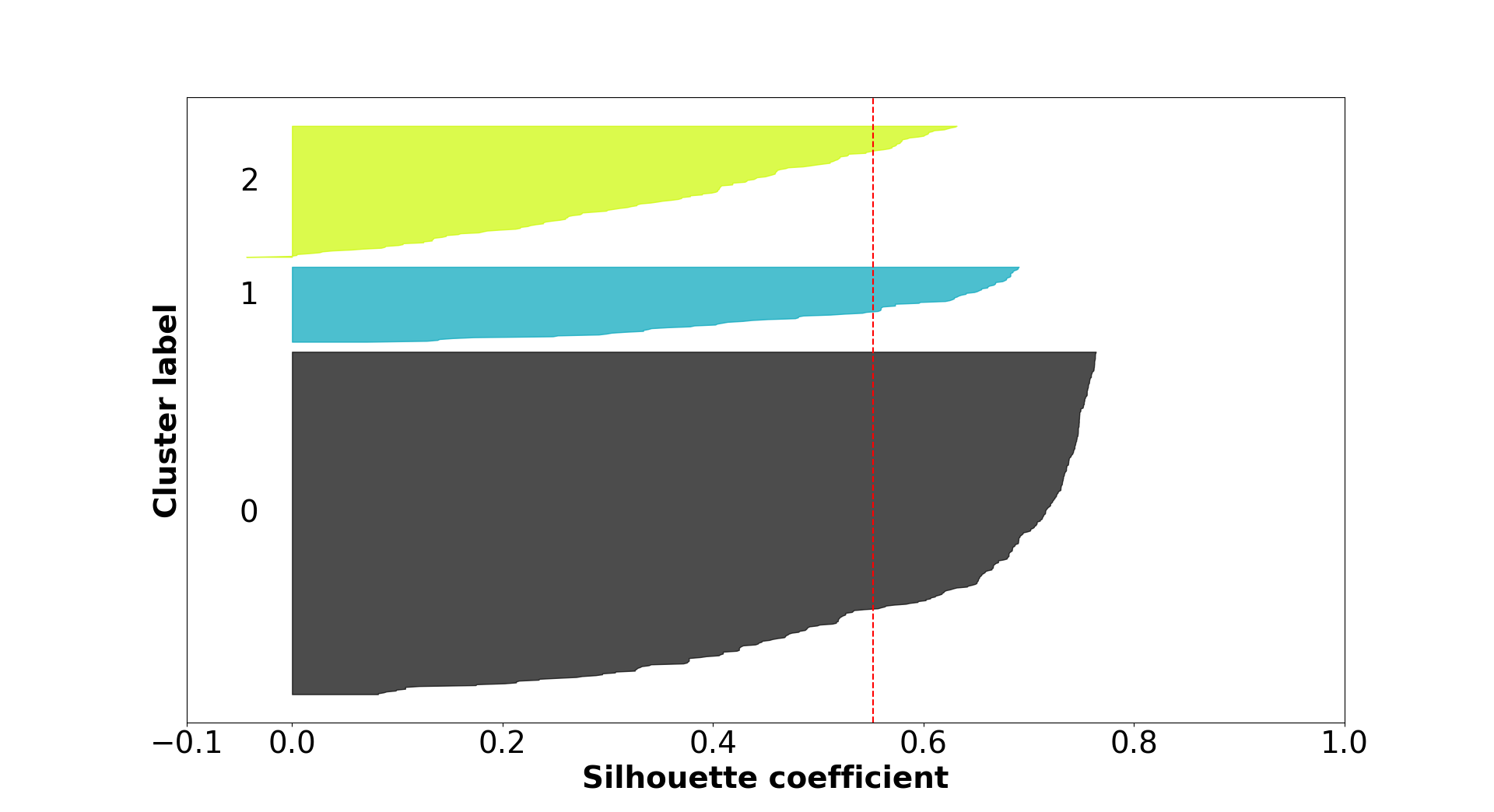}{0.5\textwidth}{}
          }
\caption{Silhouette analysis for the 4D ($tr_{\rm 1}$, $td_{\rm 1}$, $tr_{\rm 2}$, $td_{\rm 2}$; {\it upper panels}) and 6D 
($tr_{\rm 1}$, $td_{\rm 1}$, $tr_{\rm 2}$, $td_{\rm 2}$, $tr_{\rm 3}$, $td_{\rm 3}$; {\it lower panels}) clustering done on the
CSM--I/CSM--II/MAG dataset.
The k=2 results are shown in the left column and the $k =$~3 results in the right column for both cases.
\label{Fig:silhouette}}
\end{figure*}

\begin{figure*}
\gridline{\fig{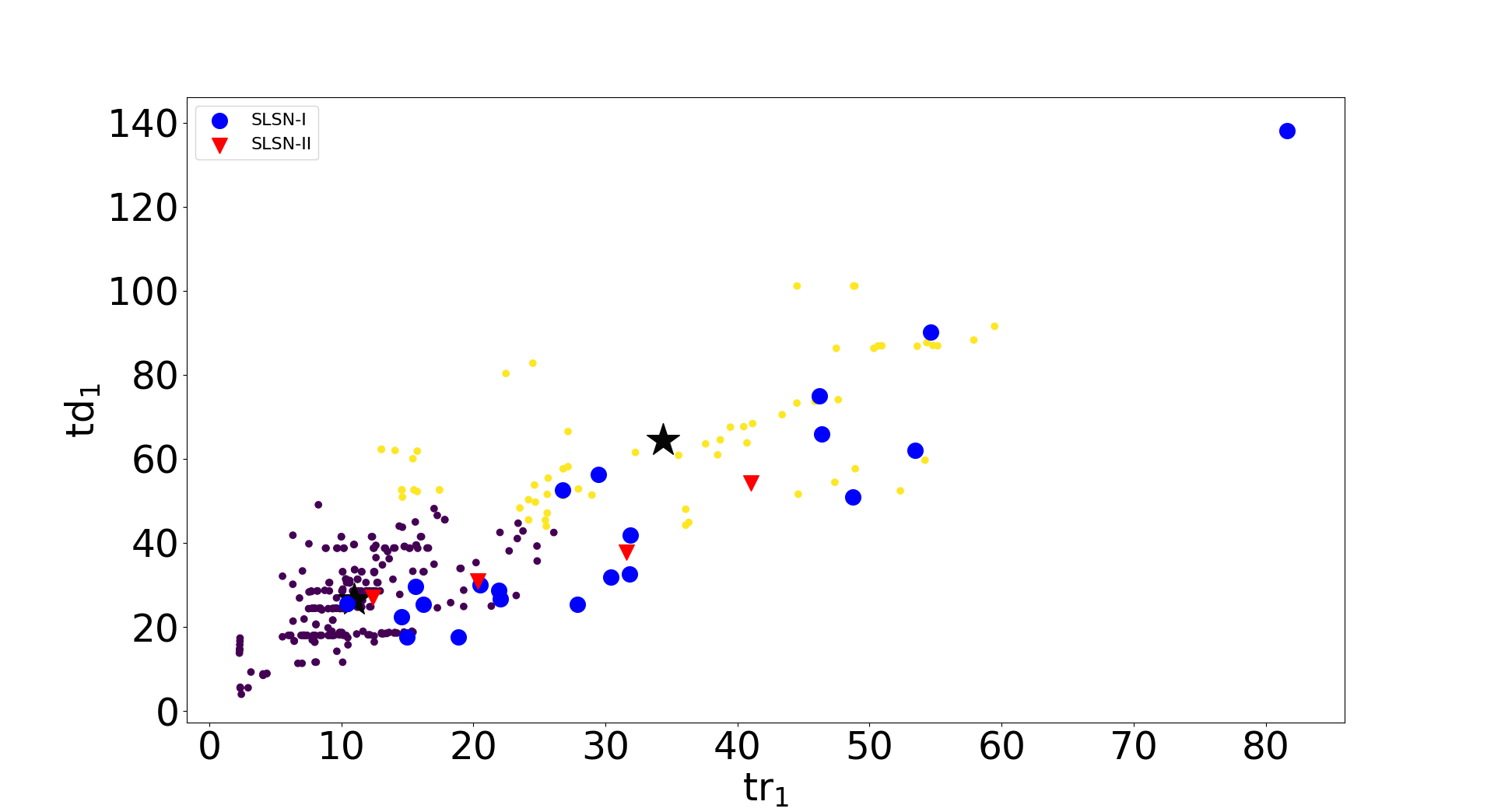}{0.5\textwidth}{}
          \fig{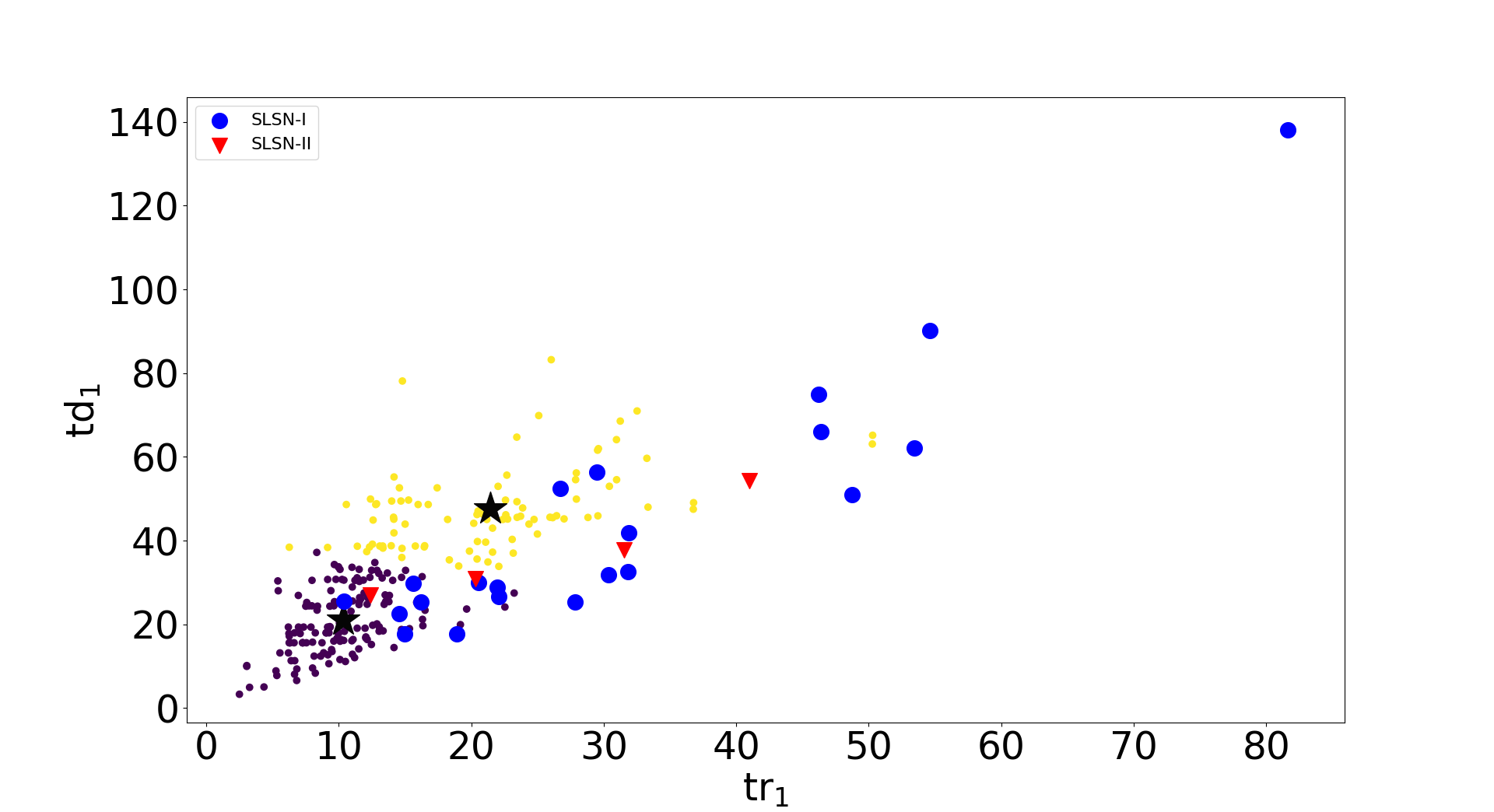}{0.5\textwidth}{}
          }
\gridline{\fig{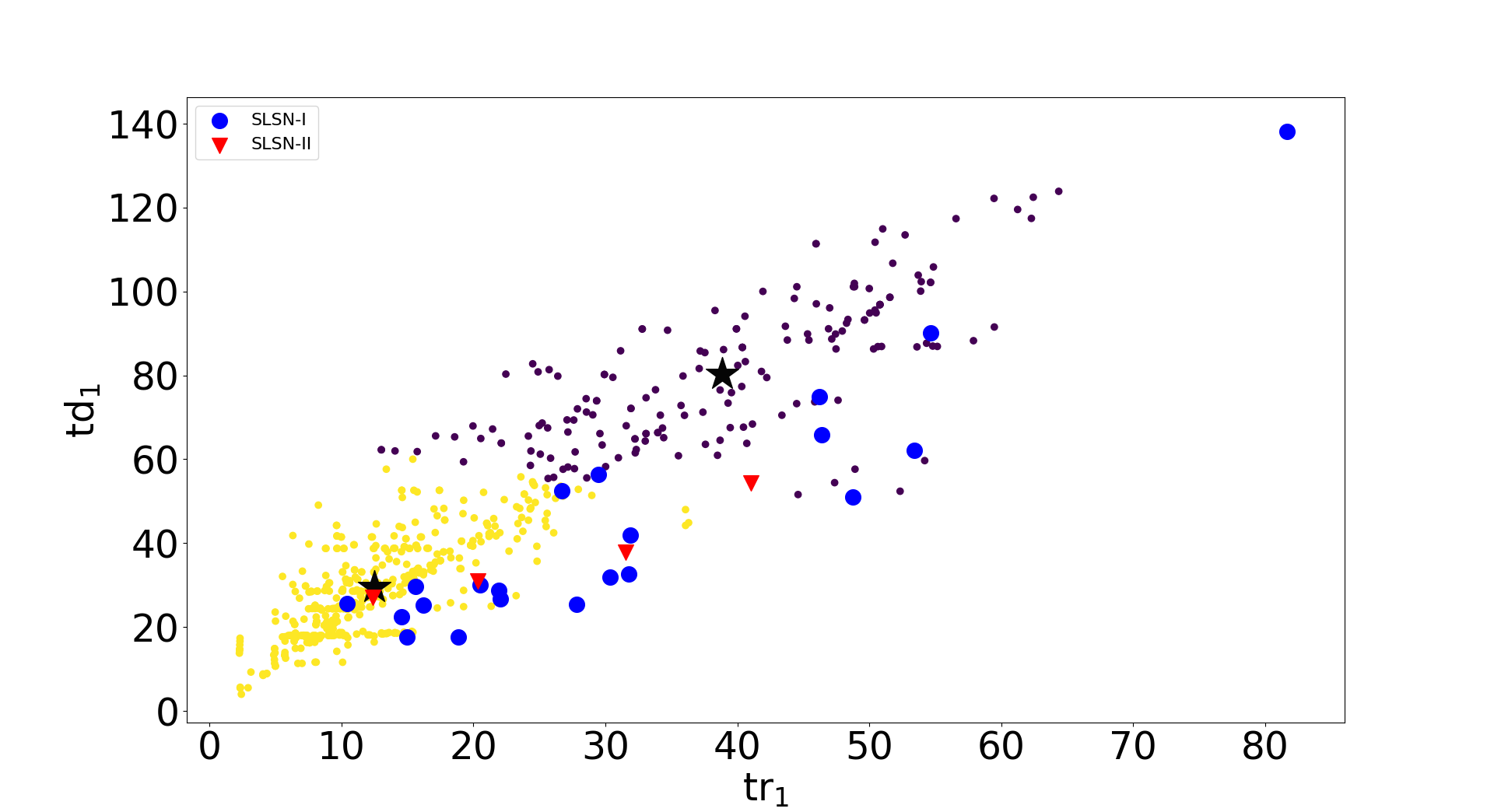}{0.5\textwidth}{}
         \fig{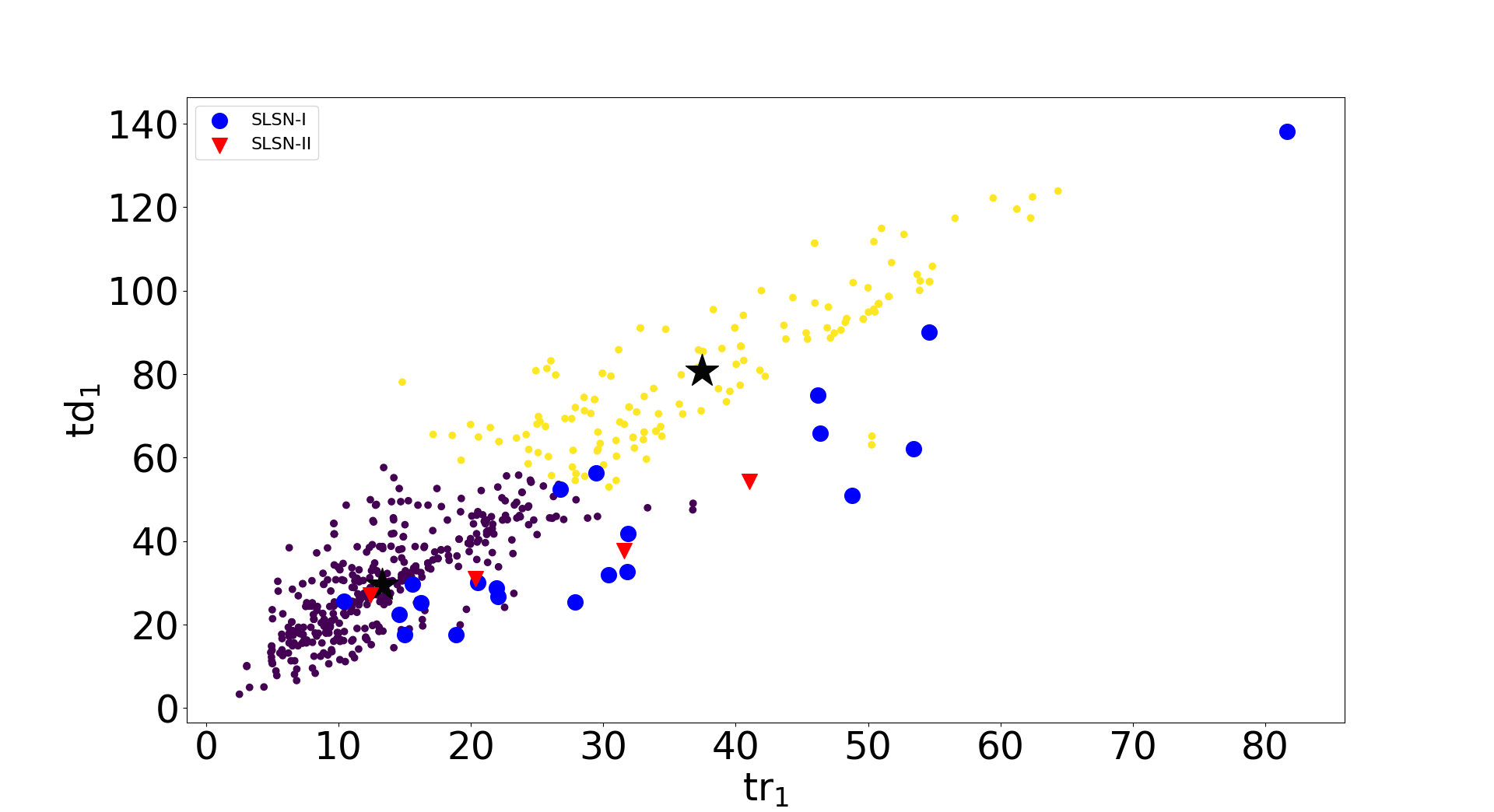}{0.5\textwidth}{}
          }
\caption{Clustering ($k =$~2) for the 2D CSM--I/CSM--II dataset ({\it upper left panel}), the 2D CSM--I$\kappa$/CSM--II$\kappa$ dataset
({\it upper right panel}), the 2D CSM--I/CSM--II/MAG dataset ({\it lower left panel}) and the 2D CSM--I$\kappa$/CSM--II$\kappa$/MAG
dataset ({\it lower right panel}). In each panel, the star symbols correspond to the cluster centroids, the blue circles to the observed
SLSN--I sample and the red triangles to the observed SLSN--II sample. For the 2D ($tr_{\rm 1}$, $td_{\rm 1}$) case, 
$k$--Means clustering is unable to find clusters that significantly overlap with the MAG and CSM models (see \ref{cluster}).
\label{Fig:cluster2D}}
\end{figure*}

\begin{figure}
\begin{center}
\includegraphics[angle=0,width=9cm]{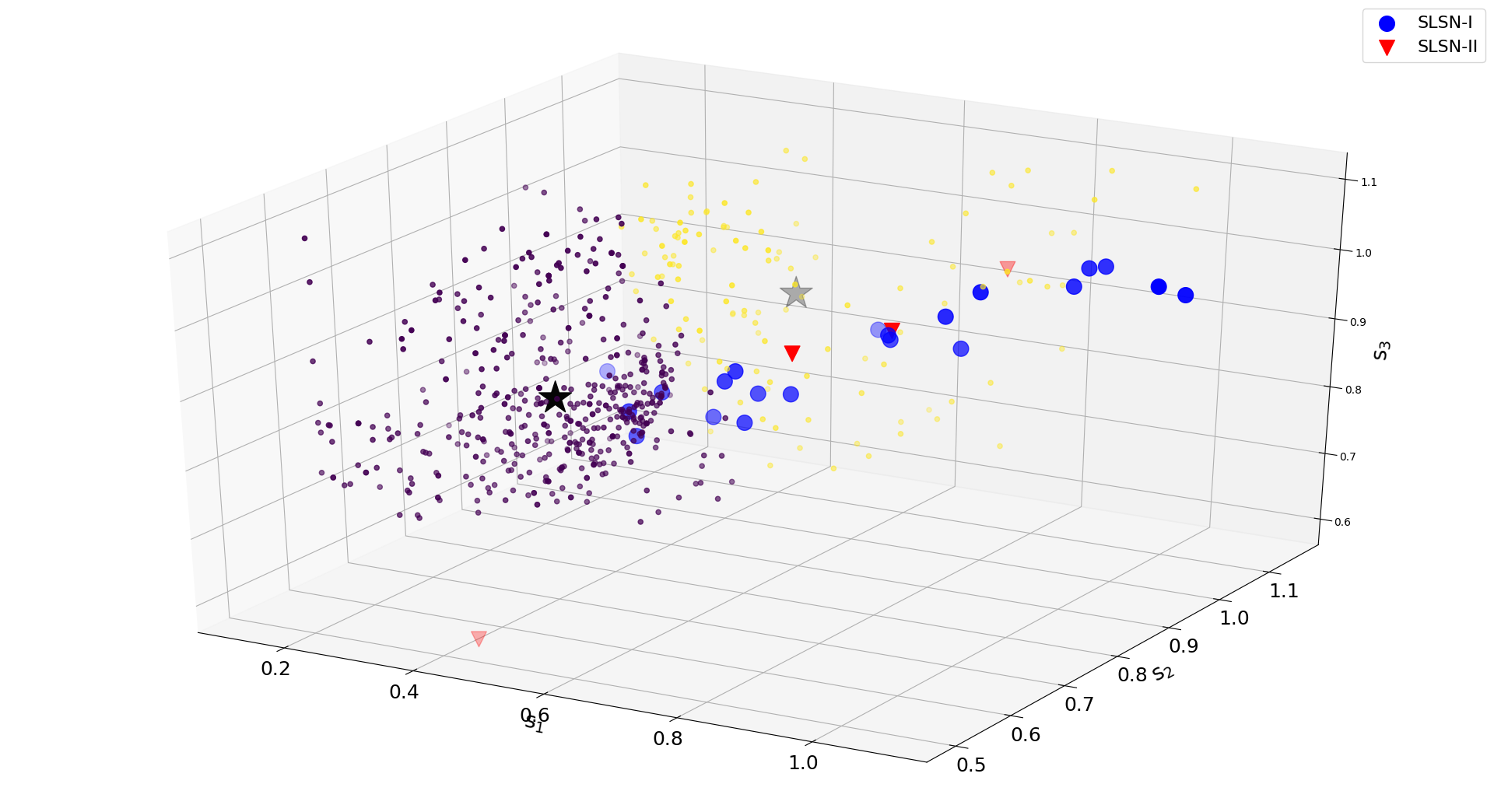}
\caption{Same as in Figure~\ref{Fig:cluster2D} but for the 3D ($s_{\rm 1}$, $s_{\rm 2}$, $s_{\rm 3}$) CSM--I/CSM--II/MAG dataset.
The computed clusters associate with the underlying model categories better than in the 2D case (see \ref{cluster}).}
\label{Fig:cluster3D}
\end{center}
\end{figure}

\section{$k$--Means Clustering Analysis}\label{cluster}

$k$--means clustering is a powerful machine learning algorithm used to categorize data via an iterative method \citep{cluster1,cluster2}. The standard
version of this algorithm finds the locations and boundaries of ``clusters'' of data by repeatedly minimizing their Euclidian distances from cluster centroids. The user can
either input the number of clusters, $k$, based on some assumption about the nature of the data, or can use a density--based (``DBSCAN'') 
approach \citep{Ester} to determine the optimal number of clusters. While $k$--means assumes clusters separated by straight--line boundaries, there exist
clustering algorithms that relax that criterion. For the scope of this work to quantitatively characterize the LC shape properties of CSM and MAG models, 
and determine if they occupy distinct areas of the parameter space, we employ $k$--means clustering analysis. 
More specifically, we use the {\it Python scikit--learn} ({\tt sklearn}) package.

$k$--means clustering analysis is often used in astronomical applications aiming to classify astronomical objects in transient search
projects \citep{2001AAS...19913004W,2004A&A...422.1113Z,2014A&A...565A..53O}. Recently, it was utilized to classify the properties of SLSNe, based
on both LC and spectroscopic features, showcasing the importance it holds for the future of the field. \citet{2018arXiv180800510N} presented their work
on $k$--means clustering analysis of SLSN nebular spectra properties. \citet{2018ApJ...854..175I} illustrated how the method can be used to identify
SLSN--I and probe their observed diversity and identified two distinct groups: ``fast'' and ``slow'' SLSN--I depending on the evolution of the LC
and the implied spectroscopic velocities and SN ejecta velocity gradients.

In this work, we use $k$--means clustering to investigate if the SLSN LC shape properties implied by 
different power input models (MAG, CSM--I and CSM--II) concentrate in distinct clusters. This may allow us to associate
observed SLSNe with proposed power input mechanisms based only on the LC properties and thus provide
a framework for SLSN characterization for future, big data transient searches like {\it LSST}. To do so, we
focus on different combinations of $k$ values and LC parameter space dimensionality ($N_{\rm D}$). Given our
prior knowledge that we are using LC shape parameter data from two categories (CSM, MAG) of models we
focus on two cases: $k =$~2 (CSM models of both I and II type and MAG) and $k =$~3 (distinct CSM--I, CSM--II and MAG models).
We also look at different values for $N_{\rm D}$: 2D datasets focusing on the primary LC timescales ($tr_{\rm 1}$, $td_{\rm 1}$),
3D datasets focusing on the LC symmetry parameters ($s_{\rm 1}$, $s_{\rm 2}$, $s_{\rm 3}$), 4D datasets focusing on the
primary and secondary LC timescales ($tr_{\rm 1}$, $td_{\rm 1}$, $tr_{\rm 2}$, $td_{\rm 2}$) and 6D datasets focusing on
the primary, secondary and tertiary LC timescales ($tr_{\rm 1}$, $td_{\rm 1}$, $tr_{\rm 2}$, $td_{\rm 2}$, $tr_{\rm 3}$, $td_{\rm 3}$) thus
covering all the LC shape parameters defined in this work (since given the 6 timescales the symmetry parameters can be constrained).
Although we only opted to perform clustering analysis for $k =$~2,3 based on prior knowledge of the number of models used in the
datasets, we also estimated the optimal number of clusters in all cases using the ``elbow'' method \citep{elbow}. This method is based
on plotting the normalized squared error of clustering ($E_{\rm N}$, defined in the next paragraph) as a function of $k$ and finding
the value of $k$ that corresponds to the sharpest gradient. This test confirmed that the optimal number of clusters for all datasets is $k =$~2.

While for the 2D and the 3D clustering we can provide visual representations of the clusters, that is impossible for the 4D and
the 6D cases. For this reason, and in order to quantify the quality and accuracy of our clustering results, we use silhouette
analysis \citep{silhouette}. Silhouette analysis yields a mean silhouette score, $\bar{S}$, and silhouette diagrams that visualize the sizes of the 
individual clusters and the $S$ score distribution of the individual data within each cluster. Negative values of $S$ correspond
to falsely classified data while values closer to unity indicate stronger cluster association. Silhouette diagrams with clusters
of comparable width and with $S$ values above the mean are indicative of accurate clustering. An example
silhouette diagram for the $k =$~2, 3 and $N_{\rm D} =$~4 case we study in this work is shown in Figure~\ref{Fig:silhouette}.
Figures~\ref{Fig:cluster2D} and~\ref{Fig:cluster3D} show the distribution of the computed clusters in the $N_{\rm D} =$~2 and 
$N_{\rm D} =$~3 cases for $k =$~2 with the SLSN--I/SLSN--II observations overplotted for comparison. 
The cluster centroids are also marked with black star symbols.
Table~\ref{T8} presents the results of clustering analysis for each $k$--$N_{\rm D}$ combination that we investigated including 
normalized classification error ($E_{\rm N}$; the square--root of the sum of squared distances of samples to their closest cluster 
center divided by the cluster size) and $\hat{S}$ as well as the computed cluster compositions (percentage of CSM--I/CSM--II and
MAG models within each cluster) and observed SLSN--I/SLSN--II cluster associations. 

\section{Results}\label{results}

\subsection{$N_{\rm D} =$~2}\label{nd2}

Our clustering analysis on the primary LC timescales ($tr_{\rm 1}$, $td_{\rm 1}$) reveals a clear dichotomy between H--rich
and H--poor CSM models in the CSM--I/CSM--II case where the first cluster ($C_{\rm 0}$)
is composed by CSM--I (and, respectively CSM--I$\kappa$) models by almost 100\%. The observed SLSN--I and SLSN--II sample
is not clearly associated with either cluster in the CSM--I/CSM--II case. For all combinations of model datasets
and values of $k$ we find the $k =$~2 choice to correspond to more accurate clustering (higher $\bar{S}$ scores). This is indicative
that the value $k =$~2 may be optimal in distinguishing between CSM--type models of either type against MAG models.
The CSM--I/CSM--II/MAG, $k =$~2 case has the highest $\bar{S}$ score and yields the first cluster ($C_{\rm 0}$) dominated
by MAG models ($\sim$~76\% of the cluster data) and the second cluster ($C_{\rm 1}$) dominated by CSM--I/CSM--II models 
($\sim$~60\% of the cluster data). Nearly $\sim$~75\% of observed SLSN--I/SLSN--II are associated with $C_{\rm 1}$ implying that,
practically, both CSM and MAG type models can reproduce SLSN LCs in terms of the primary LC timescales. As such, the
$N_{\rm D} =$~2 case does not represent a robust way to distinguish between SLSN powered by the CSM or the MAG mechanism.

\subsection{$N_{\rm D} =$~3}\label{nd3}

In this case we explore clustering for the three main LC symmetry parameters as defined in Section~\ref{lcshape}. As can
be seen in Tables~\ref{T8} the $k =$~2 cases have, in general, better $\bar{S}$ scores than the $k =$~3 cases. Another
interesting outcome is the very low normalized mean error ($<$~0.01) for all cases suggesting that clustering based on
the [$s_{\rm 1}$, $s_{\rm 2}$, $s_{\rm 3}$] dataset yields denser, more concentrated clusters around the computed centroids.

Rergardless, the most important result in this case is the strong association of observed SLSN symmetries with $C_{\rm 1}$:
$\sim$~75--76\% of SLSN--I and SLSN--II are associated with $C_{\rm 1}$ in the CSM--I/CSM--II/MAG, $k =$~2 case. In addition,
$C_{\rm 1}$ is almost entirely composed of CSM models ($\sim$~98\%). This strengthens our previous suggestion (Section~\ref{tigerfit})
that CSM models are superior to MAG models in reproducing the observed SLSN LC symmetry properties including some
fully symmetric LCs. The same result holds in the CSM--I$\kappa$/CSM--II$\kappa$/MAG, $k =$~2 case with more than half
of observed SLSN LCs associated with the cluster that is mostly composed of CSM models. This result appears to hold up in the
$k =$~3 cases. Overall, CSM and MAG models appear to be clearly distinguished in terms of LC symmetry properties 
(Figure~\ref{Fig:s1s2s3fit}).
{\it This indicates that LC shape symmetry may be critical in identifying the power input mechanism associated with observed
SLSNe, based only on photometry.}

\subsection{$N_{\rm D} =$~4}\label{nd4}

In this case we investigate $k$--means clustering for the primary and the secondary rise and decline timescales. We elect to focus
on the $k =$~2 cases since, again, they yield higher $\bar{S}$ scores. Clear distinction is recovered between H--poor and H--rich
CSM models in the CSM--I/CSM--II and the CSM--I$\kappa$/CSM--II$\kappa$ cases: $\sim$~100\% of H--poor CSM models
constitute the $C_{\rm 1}$ data in the CSM--I/CSM--II case and $\sim$~89\% of H--poor CSM models constitute
the $C_{\rm 0}$ data in the CSM--I$\kappa$/CSM--II$\kappa$ case.

For the CSM--I/CSM--II/MAG dataset we recover a cluster that
is mostly composed of CSM--type models ($C_{\rm 1}$; 60\% CSM--I/CSM--II models and 40\% MAG models) and a cluster
that is dominated by MAG models ($C_{\rm 0}$; $\sim$~20\% CSM--I/CSM--II models and $\sim$~80\% MAG models). The
majority ($\sim$~66--75\%) of SLSN--I/SLSN--II are associated with $C_{\rm 1}$ indicating preference toward CSM models yet
the correlation is not as strong as in the $N_{\rm D} =$~3 case.

\subsection{$N_{\rm D} =$~6}\label{nd6}

The last clustering analysis was performed on a six--dimensional dataset comprised of the primary, secondary and tertiary rise
and decline timescales. This is the most complete LC shape parameter dataset we investigate since it encapsulates
the three LC symmetry values, uniquely defined by their corresponding timescales. Furthermore, the use of all relevant LC shape
parameters yields the highest $\bar{S}$ scores ($\sim$~0.8 in some cases) compared to the lower--dimensionality cases.
As with all other cases, we observe that $k =$~2 clustering leads to more accurate classification therefore we only focus on these
results for our discussion.

Our results are consistent with those of the $N_{\rm D} =$~4 case yielding a cluster dominated by CSM--type models (60\%)
and a cluster dominated by MAG models ($\sim$~80\%) with the majority of SLSN--I/SLSN--II associated with the former
in the CSM--I/CSM--II/MAG case.
In particular, $\sim$~~66--75\% of observed SLSN LCs are associated with the CSM--dominated cluster.

In summary, we find that clustering of LC shape properties generally favors the CSM power input mechanism yet the MAG
mechanism cannot be ruled out. While clustering on LC timescales supports this result, it is even more robust in clustering
of LC symmetry parameters. 

\section{Discussion}\label{disc}

In this paper we explored how high--cadence photometric observations of SLSNe detected shortly after explosion can be used to 
charactize their power input mechanism. In particular, we constrained the LC shape properties of a set of
observed SLSN--I and SLSN--II focusing only on events with complete photometric coverage and searched for possible 
correlations with semi--analytic model LC shapes assuming either a magnetar spin--down (MAG) or a 
SN ejecta--circumstellar interaction (CSM) power input \citep{2012ApJ...746..121C,2013ApJ...773...76C}. 

We reiterated that there is a number of simplifying assumptions in using these semi--analytical models 
including issues with the approximation of centrally--located heating sources and homologous expansion in cases like shock 
heating where the power input can occur close to the photosphere, the assumption of constant opacity and model parameter degeneracy 
\citep{2013ApJ...773...76C,2013MNRAS.428.1020M,2018arXiv181206522K}. In addition, models predict bolometric LCs while the observed,
rest--frame SLSN LCs are pseudo--bolometric LCs computed by fitting the SED of each event based on available observations in different filters.
Regardless of all these caveats, semi--analytic models still constitute a powerful tool to study SLSNe, providing us with the potential
to investigate LC shape properties across the associated parameter space for each power input by computing a large number of models.
Nevertheless, we have supplemented our study with datasets of numerical MAG and CSM model SLSN LCs available in the literature.

To quantitatively determine whether the main proposed SLSN power input mechanisms yield model LCs with different shape
properties (rise and decline timesales and symmetry around peak luminosity) we applied $k$--means clustering analysis
for different combinations of parameters and model datasets and computed cluster associations for the observed SLSN sample. 
We highlight the main results of our analysis below:

\begin{itemize}
\item{SLSN exhibit a strong correlation between their primary rise ($tr_{\rm 1}$) and decline ($td_{\rm 1}$ timescales. Although this 
correlation is reproduced by both MAG and CSM power input models, the larger scatter found in CSM models overlaps better with the SLSN--I/SLSN--II data.}
\item{CSM models generally correspond to faster evolving LCs in agreement with observations of some SLSN--I.}
\item{MAG models fail to produce fully symmetric LCs around peak luminosity. In particular, MAG models are never found to be symmetric around the
first luminosity threshold ($s_{\rm 1, max} =$~0.54), including in cases of high gamma--ray leakage.}
\item{While the majority of CSM models also fail to produce fully symmetric LC shapes, there is a small fraction of them that do. This is in consistent
with $\sim$~24\% of SLSN--I LCs in our sample that are measured to be fully symmetric.}
\item{Symmetric SLSN LCs favor a truncated power input source that leads to faster LC decline rates past peak luminosity. The CSM model
naturally provides such a framework since forward and reverse shock power inputs are terminated. An alternative truncated input could
be energy release by fallback accretion.}
\item{MAG models fail to produce LCs with positive second derivative during the early rise to peak luminosity (concave--up). CSM models
can produce both concave--up and concave--down LCs.}
\item{$k$--means clustering analysis suggests that most observed SLSN LCs are associated with CSM power input yet the MAG model
cannot be ruled out. A multiple formation channel is therefore possible for SLSNe of both spectroscopic types.}
\item{The most distinct clustering between MAG and CSM data is found in the 3D LC symmetry parameter space ($s_{\rm 1}$, $s_{\rm 2}$, $s_{\rm 3}$). In
this case, the majority ($>$~75\%) of SLSNe are strongly associated with the CSM--dominated cluster.}
\item{LC symmetry properties, together with the shape of the LC at early times, may be key in distinguishing between different power input mechanisms
in SLSNe.}
\end{itemize}

Our results illustrate the importance of early detection and high--cadence multi--band photometric follow--up in determining the nature of SLSNe. 
As transient search surveys like {\it LSST}, {\it ZTF} and {\it Pan--STARRS} usher the new era of big data transient astronomy, a larger number of
well--constrained SLSN LCs will become available providing the opportunity to use photometry to characterize their power input mechanisms. 
This is of critical importance in the study of luminous and uncharacteristic transients in general, since photometry will be more readily available
that spectroscopy in most cases. 

We have shown that machine learning approaches like $k$--means clustering can be instrumental in helping us
characterize SLSNe based on their LC properties, namely rise and decline timescales and LC symmetry. 
This is made possible by comparing against the LC shape properties of different power input mechanisms 
using semi--analytic or numerical models. As such, it is of great importance to enhance our numerical modeling
efforts for all proposed power input mechanisms and survey a large fraction of the model parameter space.
In addition to aiding with SLSN and luminous transient characterization and classification, this will
provide us with constrains on the physical domains that enable these extraordinary stellar explosions.

\acknowledgments

We would like to thank Edward L. Robinson and J. Craig Wheeler for useful discussions and comments.
We would also like to thank our anonymous referee for suggestions and comments that improved the
quality and presentation of our paper.
EC would like to thank the Louisiana State University College of Science and the Department of Physics 
\& Astronomy for their support.

\software
{\tt Matplotlib} \citep{Hunter2007}, {\tt numpy} \citep{oliphant}, {\tt SciPy} \citep{scipy}, {\tt Scikit--learn} \citep{scikit-learn},
{\tt SuperBol} \citep{2018RNAAS...2d.230N}.

\bibliography{refs}



\end{document}